%% file: main.tex
\documentclass[acmsmall, screen]{acmart}

\usepackage{graphicx}
\usepackage{textcomp}
\usepackage{color,xcolor}
\usepackage{tcolorbox}
\usepackage{verbatim}
\usepackage{listings}
\usepackage{multirow}
\usepackage{makecell}
\usepackage{multicol}
\usepackage{diagbox}
\usepackage{subfig}
\usepackage{hyperref}
\usepackage{pifont}
\usepackage{rotating}
\usepackage{url}
\usepackage{xspace}
\usepackage{color}
\usepackage{indentfirst}

\usepackage{soul}
\renewcommand*{\hl}{}
\soulregister{\cite}7 
\soulregister{\ref}7 
\soulregister{\textbf}7 
\soulregister{\emph}7 
\soulregister{\pageref}7 

\newcolumntype{M}[1]{>{\centering\arraybackslash}m{#1}}

\usepackage[normalem]{ulem}

\usepackage[normalem]{ulem}

\AtBeginDocument{%
  \providecommand\BibTeX{{%
    \normalfont B\kern-0.5em{\scshape i\kern-0.25em b}\kern-0.8em\TeX}}}

\setcopyright{acmcopyright}
\copyrightyear{2023}
\acmYear{2023}
\acmDOI{}
\acmJournal{CSUR}
\acmVolume{}
\acmNumber{}
\acmArticle{}
\acmMonth{12}

\begin{document}

\title{Machine Learning for Actionable Warning Identification: A Comprehensive Survey}

\author{Xiuting Ge}
\orcid{0000-0003-3683-7374}
\affiliation{%
  \institution{State Key Laboratory for Novel Software Technology, Nanjing University}
  \streetaddress{No.22 Hankou Road}
  \city{Nanjing}
  \state{Jiangsu}
  \country{China}
  \postcode{210093}
}
\email{dg20320002@smail.nju.edu.cn}

\author{Chunrong Fang}
\authornote{Chunrong Fang and Zhenyu Chen are the corresponding authors.}
\orcid{0000-0002-9930-7111}
\affiliation{%
  \institution{State Key Laboratory for Novel Software Technology, Nanjing University}
  \streetaddress{No.22 Hankou Road}
  \city{Nanjing}
  \state{Jiangsu}
  \country{China}
  \postcode{210093}
}
\email{fangchunrong@nju.edu.cn}

\author{Xuanye Li}
\orcid{0009-0008-4580-2046}
\affiliation{%
  \institution{State Key Laboratory for Novel Software Technology, Nanjing University}
  \streetaddress{No.22 Hankou Road}
  \city{Nanjing}
  \state{Jiangsu}
  \country{China}
  \postcode{210093}
}
\email{1525135604@qq.com}

\author{Weisong Sun}
\orcid{0000-0001-9236-8264}
\affiliation{%
 \institution{College of Computing and Data Science, Nanyang Technological University}
  \streetaddress{Nanyang Avenue 50}
  \country{Singapore}
  }
\email{weisong.sun@ntu.edu.sg}

\author{Daoyuan Wu}
\orcid{0000-0002-3752-0718}
\affiliation{%
 \institution{The Hong Kong University of Science and Technology}
  \streetaddress{Clear Water Bay}
  \city{Hong Kong SAR}
  \country{China}
  }
\email{daoyuan@cse.ust.hk}

\author{Juan Zhai}
\orcid{0000-0001-5017-8016}
\affiliation{%
 \institution{Manning College of Information \& Computer Sciences, University of Massachusetts}
  \streetaddress{140 Governors Drive}
  \city{Amherst}
  \state{Miami}
  \country{USA}
  }
\email{juanzhai@umass.edu}

\author{Shangwei Lin}
\orcid{0000-0002-9726-3434}
\affiliation{%
 \institution{College of Computing and Data Science, Nanyang Technological University}
  \streetaddress{Nanyang Avenue 50}
  \country{Singapore}
  }
\email{shang-wei.lin@ntu.edu.sg}

\author{Zhihong Zhao}
\orcid{0009-0009-8850-5583}
\affiliation{%
  \institution{State Key Laboratory for Novel Software Technology, Nanjing University}
  \streetaddress{No.22 Hankou Road}
  \city{Nanjing}
  \state{Jiangsu}
  \country{China}
  \postcode{210093}
}
\email{zhaozhih@nju.edu.cn}

\author{Yang Liu}
\orcid{0000-0001-7300-9215}
\affiliation{%
 \institution{College of Computing and Data Science, Nanyang Technological University}
  \streetaddress{Nanyang Avenue 50}
  \country{Singapore}
  }
\email{yangliu@ntu.edu.sg}

\author{Zhenyu Chen}
\authornotemark[1]
\orcid{0000-0002-9592-7022}
\affiliation{%
  \institution{State Key Laboratory for Novel Software Technology, Nanjing University}
  \streetaddress{No.22 Hankou Road}
  \city{Nanjing}
  \state{Jiangsu}
  \country{China}
  \postcode{210093}
}
\email{zychen@nju.edu.cn}

\renewcommand{\shortauthors}{Ge et al.}

\input{src/0abstract}



\maketitle

\input{src/1introduction}
\input{src/3researchmethod}

\input{src/2backgroundandrelatedwork}

\input{src/4result}

\input{src/5discussion}
\input{src/6threat}

\input{src/7conclusion}

\bibliographystyle{ACM-Reference-Format}
\bibliography{ref}

\end{document}

%% file: src/0abstract.tex
\begin{abstract}
Actionable Warning Identification (AWI) plays a crucial role in improving the usability of static code analyzers.
With recent advances in Machine Learning (ML), various approaches have been proposed to incorporate ML techniques into AWI. These ML-based AWI approaches, benefiting from ML's strong ability to learn subtle and previously unseen patterns from historical data, have demonstrated superior performance. 
However, a comprehensive overview of these approaches is missing, which could hinder researchers \hl{and} practitioners from understanding the current process and discovering potential for future improvement in the ML-based AWI community. 
In this paper, we systematically review the state-of-the-art ML-based AWI approaches. 
First, we employ a meticulous survey methodology and gather 51 primary studies from 2000/01/01 to 2023/09/01. 
Then, we outline a typical ML-based AWI workflow, including warning dataset preparation, preprocessing, AWI model construction, and evaluation stages. In such a workflow, we categorize ML-based AWI approaches based on the warning output format. Besides, we analyze the key techniques used in each stage, along with their strengths, weaknesses, and distribution. 
Finally, we provide practical research directions for future ML-based AWI approaches, focusing on aspects like data improvement (e.g., enhancing the warning labeling strategy) and model exploration (e.g., exploring large language models for AWI).
\end{abstract}

%% file: src/1introduction.tex
\section{Introduction} \label{intro}
Static Code Analyzers (SCAs) can automatically detect defects without executing the program and have \hl{been} proven to be important and effective in software quality assurance~\cite{precision3}. 
However, SCAs often generate an overwhelming number of warnings, most of which are unactionable (e.g., false positives)~\cite{S1, S5, miningha}.
Statistics show that there are on average 40 warnings per thousand lines of source code~\cite{guizecustome1} and 35\%$\sim$91\% of warnings from SCAs are unactionable~\cite{S22}. 
Manually partitioning warnings into actionable and unactionable ones is time-consuming~\cite{6648191} and error-prone~\cite{S9}.
As such, massive unactionable warnings and the cost of manual inspection pose significant obstacles to the practical usage of SCAs~\cite{S10}.

There has been extensive research on improving the usability of SCAs. 
Various approaches~\cite{andreasen2017systematic} have been proposed to increase the precision of SCAs from the vendor's perspective, thereby minimizing false positives.
However, due to the undecidable nature of program behaviors, it is inevitable that SCAs report false positives~\cite{rice}. 
Therefore, an alternative approach is \textit{Actionable Warning Identification (AWI)}~\cite{S22, S31, S47, survey4}, which is proposed from the user’s perspective.
These AWI approaches use different techniques (e.g., clustering, ranking, pruning, automated elimination of false positives, static and dynamic combination analysis, or simplifying manual inspection)~\cite{S31} to postprocess warnings reported by SCAs, thereby classifying or \hl{ranking} actionable warnings. 
Machine Learning (ML)-based AWI approaches are notably popular, which train a model with historical warnings and use it to identify actionable warnings on new ones~\cite{S4, S18}.
Due to ML's powerful ability to learn subtle and previously unseen patterns from historical data, ML-based AWI approaches have demonstrated superior performance in enhancing the usability of SCAs~\cite{S24, S25}.

Over the past few years, the substantial progress in the ML-based AWI community has attracted considerable attention from researchers and practitioners.
Currently, several existing AWI literature reviews have been proposed to enumerate~\cite{S22} or categorize~\cite{S31, S47, survey4} specific AWI approaches.
Different from the existing reviews that mainly focus on postprocessing techniques of warnings, ML-based AWI approaches require unique characteristics (i.e., the heavy reliance on warning datasets, warning features, and model selection) to identify actionable warnings.
However, the existing reviews overlook such unique characteristics, which could present various challenges in developing new and advanced ML-based AWI approaches.
For example, warning datasets highly depend on SCAs employing different techniques and projects using various development languages, which could affect the warning feature extraction ways.
Also, different warning labeling strategies (e.g., closed warning-based heuristic \cite{S10}) could impact the quality of warning datasets and thus affect the ML-based AWI performance.
Moreover, different categories of features (e.g., content-based or sequential) necessitate selecting appropriate models for warning representation, leading to varying AWI performance.
These diverse design options could hinder researchers and practitioners from further advancements in the ML-based AWI research direction.

To fulfill the above gaps, we are the first to conduct a comprehensive survey by retrospectively examining the current state-of-the-art ML-based AWI studies after years of development. 
Through analyzing these studies, we first outline a typical ML-based AWI workflow, involving warning dataset preparation, warning dataset preprocessing, AWI model construction, and AWI model evaluation stages.
In such a workflow, we categorize ML-based AWI approaches based on the warning output format. 
Besides, we detail the techniques used in each stage of such a workflow by discussing their strengths and weaknesses and presenting their distribution across different categories of ML-based AWI approaches.
Finally, we provide several practical research directions for the ML-based AWI community.
In summary, we believe that our survey can help researchers and practitioners gain a comprehensive understanding and foster progress toward advanced practices in the ML-based AWI field.
Our survey makes the following major contributions:
\begin{itemize}

    \item \textbf{Survey methodology.} We employ a meticulous survey methodology across five digital libraries from 2000 to 2023 to gather 51 primary ML-based AWI studies. 

    \item \textbf{ML-based AWI.} We outline a typical workflow of applying ML techniques \hl{to} AWI approaches, which involves warning dataset preparation, warning dataset preprocessing, AWI model construction, and AWI model evaluation.

    \item \textbf{Elaborate study.} We conduct a detailed analysis of the typical ML-based AWI workflow. 
    Such an analysis includes the categorization of ML-based AWI approaches based on the warning output format as well as the discussion of key techniques with associated strengths, weaknesses, and distribution across different categories of ML-based AWI approaches.

    \item \textbf{Practical directions.} We highlight nine practical research directions for future ML-based AWI from the perspectives of data improvement and model exploration.
    
    \item \textbf{Available artifacts.} We share primary ML-based AWI studies along with associated artifacts in a public repository~\cite{mylink}, which facilitates following and extending our survey.
\end{itemize}
  

%% file: src/3researchmethod.tex
\section{Survey Methodology} \label{collectpaper}
Guided by the work of Budgen et al. \cite{budgen2006performing}, we collect the relevant ML-based AWI studies from the population via a well-designed survey methodology. Such a methodology includes data sources, search keywords, selection criteria, selection procedure of primary studies, and data extraction and synthesis. Further, we perform the trend observations based on the primary studies.


\textbf{Data sources.} We search five popular digital libraries, including (1) \href{https://ieeexplore.ieee.org/}{IEEE Xplore}, (2) \href{https://dl.acm.org/}{ACM}, (3) \href{https://www.sciencedirect.com/}{ScienceDirect}, (4) \href{https://link.springer.com/}{Springer Link}, and (5) \href{https://onlinelibrary.wiley.com/}{Wiley}. These libraries archive various and leading journals and conferences from the software engineering domain \cite{hoonlor2013trends}.

\textbf{Search keywords.} 
Based on the work of Muske et al. \cite{S31}, we extend the search keywords to \hl{identify} the relevant ML-based AWI studies from \hl{every} digital library. Table \ref{tab:keywords} shows the search keywords. 
Such search keywords involve four goals (i.e., ML, static analysis, warning, and identification) with a total of 19 keywords. 
To search ML-based AWI studies, the four goals are required to be incorporated to create a complete search string. 
Specifically, for each goal, the logical operator ``AND'' is used. For each keyword in each goal, the logical operator ``OR'' is used. 
With these search strings, we search the above five digital libraries by performing the keyword-based matching in the metadata (i.e., title, abstract, and keywords) of the study \cite{keele2007guidelines}.
The search scope in each digital library is from 2000/01/01$\sim$2023/09/01. 
Such a search scope is considered because ML techniques gradually gain attention in the AWI community \cite{S22}.

\begin{table}[]
    \centering
    \caption{Search keywords.}
    \label{tab:keywords}
         \scalebox{0.9}{
    \setlength{\tabcolsep}{0.2mm}{
    \begin{tabular}{|c|c|p{340pt}<{\centering}|}
    \hline
        \textbf{No.} &  \textbf{Goal} & \textbf{Keyword}\\ \hline \hline
         1 & machine learning & 1) machine learning, 2) deep learning\\ \hline
         2 & static analysis & 1) static analysis, 2) automated code analysis, 3) source code analysis \\ \hline
         3 & warning & 1) warning, 2) alert, 3) alarm, 4) violation \\ \hline
        4 & identification & 1) identifying, 2) elimination, 3) reduction, 4) pruning, 5) classification, 6) prioritization, 7) ranking, 8) reviewing, 9) inspection, 10) simplification \\ \hline
    \end{tabular}
    }
    }
\end{table}

\textbf{Selection criteria.} 
The selection criteria are used to further determine the relevant ML-based AWI studies from the search keywords-based results. 
Such criteria are initially obtained by steering 10 arbitrarily selected studies and are refined based on the pilot search results. 
Specifically, there are three inclusion criteria, including (1) studies that incorporate ML techniques into AWI; (2) studies that record the contextual items, including datasets, features, ML models, and evaluation details; and (3) studies that locate from 2000/01/01$\sim$2023/09/01. 
There are seven exclusion criteria, including (1) studies that identify warnings via non-ML techniques, e.g., \cite{oasis, li2014residual, junker2012smt}; (2) studies that improve the precision of SCAs, e.g., \cite{S43}; (3) studies that evaluate the precision of SCAs, e.g., \cite{precision3}; (4) studies that track the software quality evolution or SCA rule configuration by investigating warnings, e.g., \cite{quality1, guizecustome1}; (5) studies that identify vulnerable/malicious behaviors via ML techniques and static analysis, e.g., \cite{pereira2021machine}; (6) studies that recommend SCAs to different projects, e.g., \cite{nunes2017combining}; and (7) studies that are not peer-reviewed, e.g., \cite{S2-similar}.
When both the metadata and full text of a study satisfy inclusion and exclusion criteria, and this study is determined to be relevant.


\begin{table}[]
    \centering
    \caption{Selection results of primary studies.}
    \label{tab:summary}
     \scalebox{0.9}{
    \setlength{\tabcolsep}{0.2mm}{
    \begin{tabular}{|p{100pt}<{\centering}|p{150pt}<{\centering}|p{100pt}<{\centering}|}
        \hline
        \textbf{Digital library} & \textbf{No. of studies from the search keywords-based results} & \textbf{No. of studies after the selection criteria}\\ \hline \hline
        IEEE Xplore & 85 & 25 \\  \hline
        ACM & 913 & 13 \\  \hline
        ScienceDirect & 35 & 4  \\ \hline
        Springer Link & 70  & 5 \\ \hline
        Wiley & 150 & 0\\  \hline \hline
        All & 1253 & 47\\ \hline
        \multicolumn{2}{|c|}{Removing 15 duplicate studies} & 32 \\ \hline
        \multicolumn{2}{|c|}{Snowballing on 32 studies} & 19 \\ \hline
        \multicolumn{2}{|c|}{A total of studies} & 51 \\ \hline
    \end{tabular}
    }
    }
\end{table}

\textbf{Selection procedure of primary studies.} 
Table \ref{tab:summary} shows the detailed selection results of ML-based AWI studies. 
Specifically, the search keywords-based results contain 1253 studies. 
Then, after applying the selection criteria for each study, there involve 47 studies. 
After that, we rely on the titles of studies to remove 15 duplicate studies and determine 32 distinct studies. 
However, the search keywords in Table \ref{tab:keywords} may be incomplete due to the terminological differences among studies. 
To alleviate this problem, we conduct the snowballing \cite{wohlin2014guidelines}, including backward and forward snowballing.
Given a good start set, the snowballing can help more effectively and efficiently locate high-quality studies in obscure locations \cite{greenhalgh2005effectiveness}. 
As such, we take 32 distinct studies as a start set, conduct the snowballing on this start set, and use the selection criteria to search 19 other relevant studies. 
Finally, we determine 51 primary ML-based AWI studies. 
Particularly, in the selection procedure of studies, we conduct the 2-pass review. That is, every study goes through two authors (i.e., the first and third authors in our survey). 
When there is a disagreement on the inclusion or exclusion of a study, the two authors make the discussion and resolve the difference. 

\textbf{Data extraction and synthesis.} To ensure data extraction consistency among primary studies, we formulate a unified data extraction form based on a pilot study. 
To ensure the data extraction correctness in each primary study, the extracted data is checked by two authors. Specifically, the first author of our survey extracts the data from all primary studies. The extracted data is assigned and checked by the third author of our survey. 
If there is disagreement on the extracted data, the two authors discuss and reach an agreement. In all, 22 data items are extracted for each primary study.
Due to the limited space, the detailed data items are shown in a public repository \cite{mylink}. 
After that, the extracted data is synthesized via quantitative and qualitative analysis, which can assist us in answering research questions in Section \ref{result}.

\begin{figure}
    \centering
    \includegraphics[scale=0.35]{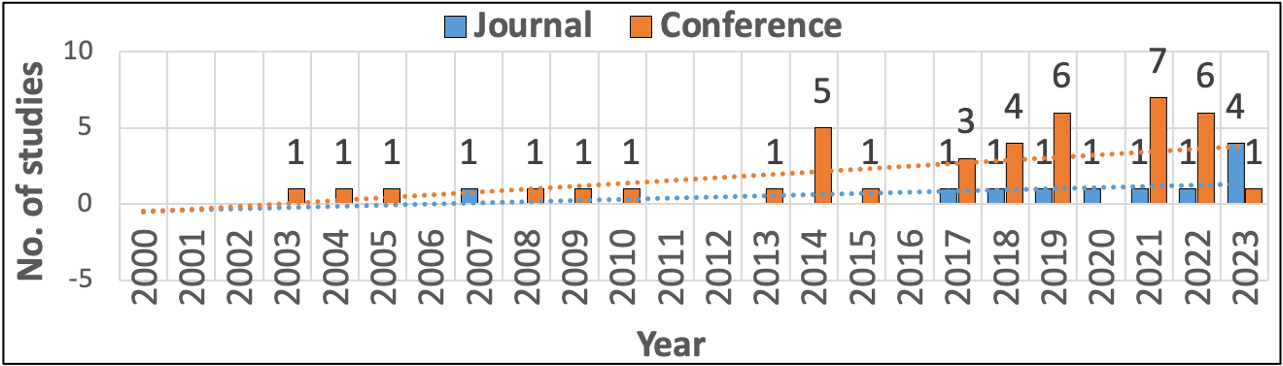}
    \caption{Distribution of primary studies. The blue (yellow) dashed lines represent the trends \hl{in} the number of journal (conference) studies over years.}
    \label{fig:disbyyear}
\end{figure}

\textbf{Trend observations.} Fig. \ref{fig:disbyyear} shows the statistical distribution of 51 primary studies. 
The results show that most studies are from the conference, which accounts for 78\% (i.e., 40) of 51 primary studies. 
The remaining studies (i.e., 11) are from the journal. 
Besides, the number of studies by year has \hl{steadily increased}, which indicates that the ML-based AWI community has received more and more attention from researchers and practitioners in recent years.

%% file: src/2backgroundandrelatedwork.tex
\section{Background and Related Work} \label{back}
\textbf{Static analysis warnings.} To help developers quickly and easily locate software defects, warnings reported by SCAs generally involve characteristics with the category, severity, message, and location. 
Of these, the warning location consists of the class and method information containing this warning and the warning line numbers.
Due to the limited space, a warning example from SpotBugs is shown in a public repository \cite{mylink}.
Based on whether warnings are \hl{ignored} by developers, warnings can be divided into actionable and unactionable ones \cite{miningha, ViolationTracker, S10}. 
\hl{Such a classification focuses on emphasizing the importance of developer perception \cite{newpaper}. 
Specifically, an actionable warning is acted on and fixed by developers. By contrast, an unactionable warning is ignored by developers due to many possible reasons (e.g., the over-approximation behaviors \cite{rice} and bugs \cite{limit, limit2} of SCAs).}
Formally, given a set of commits \emph{$C = \{c_1, ..., c_i, ..., c_n\}$} in a project (\emph{c$_n$} is the latest commit), a SCA is used to scan the source code of \emph{$c_i$} and a set of warnings \emph{$W_i = \{w_{i1}, ..., w_{ij}, ..., w_{im}\}$} (\emph{m} is the number of warnings in \emph{$c_i$}) is obtained. 
If \emph{$w_{ij}$} disappears via the warning-related source code change in any commit from \emph{$c_{i+1}$} to \emph{$c_n$}, \emph{$w_{ij}$} is denoted as an actionable warning. 
If \emph{$w_{ij}$} persists from \emph{$c_{i+1}$} to \emph{$c_n$}, \emph{$w_{ij}$} is denoted as an unactionable warning.





\textbf{AWI.} Given a set of warnings reported by a SCA, AWI \hl{\cite{S31, newpaper}} is to identify actionable warnings from all reported warnings, thereby (1) reducing the number of warnings before reporting them to SCA users; (2) prioritizing warnings that are more likely to be actionable ahead of other warnings; and (3) simplifying manual inspection effort of warnings.

\textbf{ML-based AWI.} \label{definition}
ML-based AWI is to extract features from historical warnings, learn a model on the extracted features, and use this model to identify actionable warnings from targeted warnings. 
ML-based AWI is often formalized into a supervised learning-based problem.
Given a set of historical warnings \emph{$(X, Y) = \{(x_1, y_1), ..., (x_i, y_i), ..., (x_n, y_n)\}$}, \emph{$(x_i, y_i)$} (\emph{$1 \le i \le n$}) is a historical warning.
As for \emph{$x_i \in X = \{wf_{i,1}, ..., wf_{i,j}, ..., wf_{i,m}\}$}, \emph{$wf_{i, j}$} (\emph{$1 \le j \le m$}) is a warning feature.
As for \emph{$y_i \in Y = \{0,1\}$}, \emph{$y_i$} is the warning label of \emph{$x_i$}, where \emph{$y_i = 0/1$} denotes that \emph{$x_i$} is an unactionable/actionable warning respectively.
ML-based AWI relies on historical warnings \emph{$(X, Y)$} to learn a decision function \emph{Y = f(X)}, aiming to describe the mapping relations between warning features and warning labels. 
When given a targeted warning \emph{X$_{target}$}, the learned decision function is used to predict a corresponding output \emph{y$_{target}$}. 
In general, \hl{such an output format} could be a binary or continuous value.


\textbf{Existing literature reviews.} 
Currently, there have been four related AWI literature reviews \cite{S22, S31, S47, survey4}. 
Table \ref{tab:difference} shows the differences between our survey and the four reviews. 
\emph{First}, our survey spans a longer period of search scope (i.e., about 23 years from 2000/01/01 to 2023/09/01) compared to the four reviews.  
Such \hl{an} enlarged search scope helps our survey increase about 20 new studies that \hl{have} never appeared in the four reviews, which can facilitate researchers and practitioners tracking and refreshing the current state-of-the-art ML-based AWI process. 
\emph{Second}, our survey conducts a more comprehensive analysis of ML-based AWI.
On the one hand, unlike the four reviews that enumerate or categorize AWI approaches based on the postprocessing techniques of warnings, our survey categorizes ML-based AWI approaches based on the warning output format.
Such an approach category can help researchers and practitioners gain a deeper understanding of how ML techniques work for AWI.
On the other hand, in each stage of the typical ML-based AWI workflow, our survey analyzes the key techniques with associated strengths and weaknesses as well as exhibits the distribution of these key techniques across different categories of ML-based AWI approaches.
Such a detailed analysis can provide researchers and practitioners with a thorough understanding of the ML-based AWI field. 
\emph{Third}, our survey provides more targeted guidelines for the ML-based AWI field. 
Based on the analysis results, our survey highlights practical guidelines from the perspective of data improvement and model exploration when applying ML techniques in AWI, which can assist researchers and practitioners \hl{in enhancing} ML-based AWI approaches in a targeted manner.
Particularly, the goal between our survey and the review of Guo et al. \cite{survey4} is different. Our survey focuses on AWI, while the review of Guo et al. only centers on false positive mitigation.
As shown in the above warning \hl{classification}, AWI can embrace a more extensive research scope than false positive mitigation. 
\hl{This} indicates that the findings of our survey could be more practical and flexible than those of the review of Guo et al.


%% file: src/4result.tex
\begin{table}[]
    \centering
     \caption{Comparison of differences between the existing literature reviews and our survey.}
    \label{tab:difference}
     \scalebox{0.65}{
    \setlength{\tabcolsep}{0.1mm}{
    \begin{tabular}{|c|c|p{25pt}<{\centering}|p{50pt}<{\centering}  |p{40pt}<{\centering}|p{40pt}<{\centering}|p{40pt}<{\centering}|p{40pt}<{\centering}|p{40pt}<{\centering}  |p{40pt}<{\centering}|p{40pt}<{\centering}|p{40pt}<{\centering}|p{40pt}<{\centering}|p{40pt}<{\centering}|p{40pt}<{\centering}|}
    \hline
    
\multirow{2}{*}{\textbf{Study}} &	\multirow{2}{*}{\textbf{\makecell{Search\\scope}}} &	\multirow{2}{*}{\textbf{\makecell{Appr-\\oach \\cate-\\gory}}} & \multicolumn{2}{c|}{\textbf{\makecell{Warning dataset \\preparation}}}	& 	\multicolumn{3}{c|}{\textbf{\makecell{Warning dataset \\processing}}} &			\multicolumn{4}{c|}{\textbf{AWI model construction}} &			\multicolumn{2}{c|}{\textbf{\makecell{AWI model \\evaluation}}} & \multirow{2}{*}{\textbf{\makecell{Guide-\\lines \\ in ML-\\based \\ AWI}}}
\\ \cline{4-14}
& & & \textbf{\makecell{Warning \\dataset ac-\\quisition}} &	\textbf{\makecell{Warning \\dataset \\labeling}}& \textbf{\makecell{Warning \\feature \\category}} &	\textbf{\makecell{Warning \\feature \\selec-\\tion}} &	\textbf{\makecell{Warning \\dataset \\rebalan-\\cing}}	& \textbf{\makecell{Model \\category \\in AWI}}	& \textbf{\makecell{Learn-\\ing \\category \\in AWI}}	& \textbf{\makecell{AWI \\model \\struc-\\ture}}	& \textbf{\makecell{AWI \\constru-\\ction \\scenario}}	& \textbf{\makecell{Valida-\\tion \\strategy \\in AWI}}	& \textbf{\makecell{Perfor-\\mance \\measure \\in AWI}} & \\ \hline \hline

	\cite{S22}	&	1998-2009	&	\ding{51}	&	\ding{55}	&	\ding{55}	&	\ding{51}	&	\ding{55}	&	\ding{55}	&	\ding{51}	&	\ding{55}	&	\ding{55}	&	\ding{55}	&	\ding{51}	&	\ding{51} & \ding{55}	\\ \hline
	\cite{S31}	&	2002-2006	&	\ding{51}	&	\ding{55}	&	\ding{55}	&	\ding{55}	&	\ding{55}	&	\ding{55}	&	\ding{55}	&	\ding{55}	&	\ding{55}	&	\ding{55}	&	\ding{55}	&	\ding{55}	& \ding{55} \\ \hline
	\cite{S47}	&	2002-2020	&	\ding{51}	&	\ding{55}	&	\ding{55}	&	\ding{55}	&	\ding{55}	&	\ding{55}	&	\ding{55}	&	\ding{55}	&	\ding{55}	&	\ding{55}	&	\ding{55}	&	\ding{55}	& \ding{55} \\ \hline
    \cite{survey4}	&	2003-2022	&	\ding{51}	&	\ding{51}	&	\ding{55}	&	\ding{55}	&	\ding{55}	&	\ding{55}	&	\ding{51}	&	\ding{55}	&	\ding{55}	&	\ding{55}	&	\ding{55}	&	\ding{51}	& \ding{55} \\ \hline
	Ours	&	2000-2023	&	\ding{51}	&	\ding{51}	&	\ding{51}	&	\ding{51}	&	\ding{51}	&	\ding{51}	&	\ding{51}	&	\ding{51}	&	\ding{51}	&	\ding{51}	&	\ding{51}	&	\ding{51}	& \ding{51} \\ \hline
    \end{tabular}
    }
    }
\end{table}

\section{Detailed Analysis of ML-based AWI} \label{result} 
\subsection{Typical ML-based AWI Workflow and Research Questions}
In this section, we describe a typical ML-based AWI workflow and present the proposed Research Questions (RQs) based on such a workflow.

\begin{figure}
    \centering
    \includegraphics[scale = 0.36]{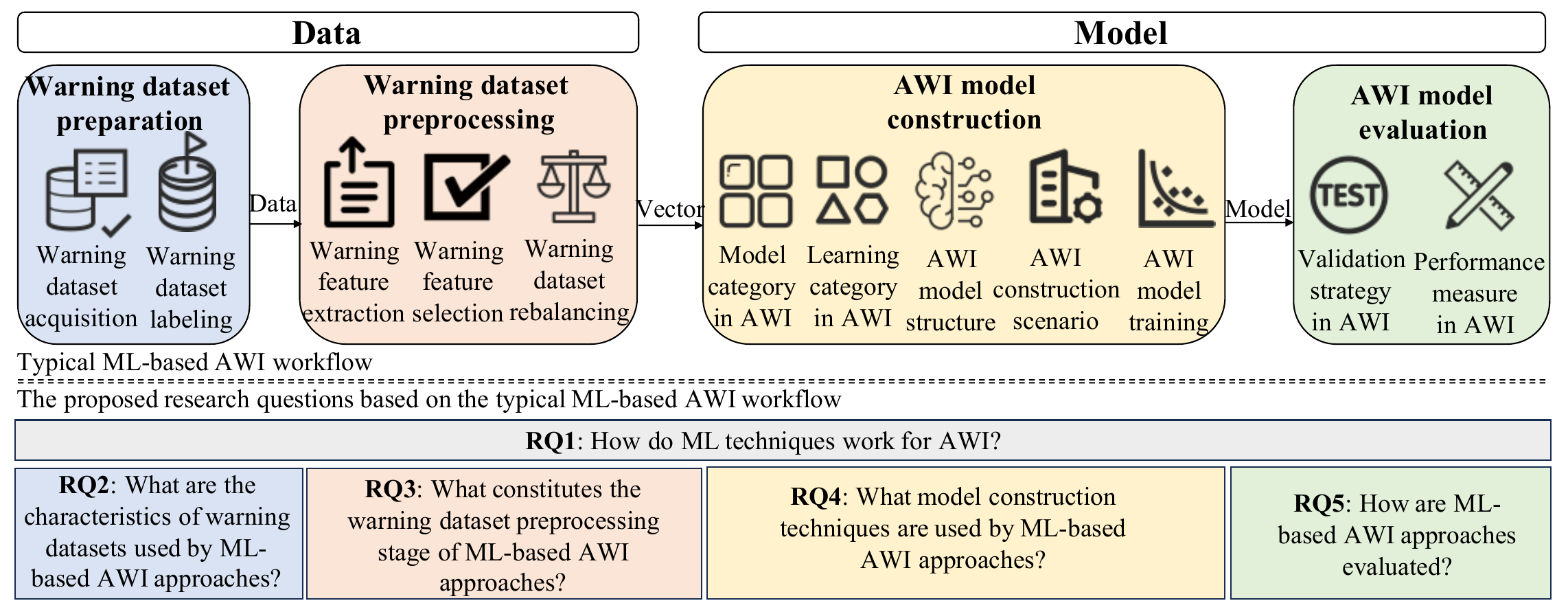}
    \caption{Typical ML-based AWI workflow and the proposed RQs based on such a workflow.}
    \label{fig:workflow}
\end{figure}

\textbf{Typical ML-based AWI workflow.} By analyzing the primary studies, we outline a typical ML-based AWI workflow.
As shown in Fig. \ref{fig:workflow}, this workflow mainly contains four stages, i.e., warning dataset preparation, warning dataset preprocessing, AWI model construction, and AWI model evaluation. 
In particular, the first and second stages focus on the data part of ML-based AWI, and the third and fourth stages focus on the model part of ML-based AWI.

\emph{(1) In the warning dataset preparation stage}, given a set of projects and SCAs, the collected warnings with associated labels are returned.
This stage mainly contains warning dataset acquisition and labeling. The warning dataset acquisition collects warnings from SCAs with various static analysis techniques and projects with different development languages. The warning dataset labeling assigns labels for collected warnings. 
In general, the well-prepared warnings are split into the training and test set.

\emph{(2) In the warning dataset preprocessing stage}, the well-prepared warnings are preprocessed via different ways.
According to the existing ML-based AWI approaches \cite{S4, S8, S24, S25}, the common ways to preprocess warnings include warning feature extraction, warning feature selection, and warning dataset rebalancing.
Specifically, the warning feature extraction mines useful features from the well-prepared warnings. 
The warning feature selection is to select the discriminative warning feature subset from the original warning features.
The warning dataset rebalancing is to alleviate the class imbalance in the well-prepared warnings. 
In general, for ML-based AWI approaches, the warning feature extraction is a necessary step, while the remaining two ways are optional steps.

\emph{(3) In the AWI model construction stage}, different ML techniques (e.g., Random Forest) are used to train an AWI model based on the training set.
Generally, this stage involves model category, learning category, AWI model structure, AWI construction scenario, and AWI model training. 
The model category (e.g., deep learning) denotes the category of ML techniques used for AWI. 
The learning category (e.g., supervised learning) represents how an ML-based AWI model learns from the training set.
The AWI model structure (e.g., the base structure) denotes how to organize ML techniques to train an AWI model.
The AWI construction scenario (e.g., within project) represents the application scenario of an ML-based AWI model.
The AWI model training is responsible for constructing an optimal ML-based AWI model from the training set.

\emph{(4) In the AWI model evaluation stage}, the performance of the well-constructed AWI model is evaluated on the test set by setting the validation strategy and selecting the performance measure. 
The validation strategy signifies how the well-prepared warnings are split into the training and test sets. The performance measure shows how a well-constructed AWI model performs.

\textbf{Research questions.} Inspired by the typical ML-based AWI workflow, we propose five Research Questions (RQs) to analyze the existing ML-based AWI studies.  
\begin{itemize}
    \item \emph{RQ1: How do ML techniques work for AWI?} 
    \item \emph{RQ2: What are the characteristics of warning datasets used by ML-based AWI approaches?} 
    \item \emph{RQ3: What constitutes the warning dataset preprocessing stage of ML-based AWI approaches?} 
    \item \emph{RQ4: What model construction techniques are used by ML-based AWI approaches?} 
    \item \emph{RQ5: How are ML-based AWI approaches evaluated?}
\end{itemize}

As described in Fig. \ref{fig:workflow}, RQ1 analyzes the primary studies from a holistic perspective of ML-based AWI, and RQ2 $\sim$ RQ5 analyze the primary studies from an atomistic perspective of ML-based AWI. 
Specifically, RQ1 aims to classify ML-based AWI approaches based on the warning output format, thereby providing an overall understanding of ML techniques' application in AWI.
RQ2 $\sim$ RQ5 correspond with the four stages of such a workflow, which aim to provide insights for the key techniques used in the warning dataset preparation, warning dataset preprocessing, AWI model construction, and AWI model evaluation, respectively.

\subsection{RQ1: How do ML techniques work for AWI?}  \label{rq1} 
As shown in Section \ref{definition}, given a targeted warning in the test set, the well-constructed ML-based AWI model in the typical ML-based AWI workflow gives an output for this warning. 
Such an output format could be a binary value, a continuous value, or both binary and continuous values. 
Based on the warning output format, ML-based AWI approaches can be divided into classification, ranking, and combination approaches. 
Further, we illustrate the three categories of approaches along with their strengths and weaknesses, thereby understanding the application of ML techniques in AWI. 

\textbf{Classification approach.} The classification approach aims to learn an ML-based AWI classifier based on historical warnings and use this classifier to classify targeted warnings into actionable and unactionable ones. 
That is, given a targeted warning, the output of this classifier is a binary value, which indicates an actionable or unactionable one. 
\hl{Subsequently,} actionable warnings are shown to developers for inspection, while unactionable warnings are pruned. 
\hl{Thus}, the classification approach can reduce the number of warnings for manual inspection.
However, since the pruned warnings are not guaranteed to be false positives, the classification approach may result in false negatives. 
In the primary studies, 31 studies \cite{S2, S3, S4, S6, S7, S9, S10, S12, S13, S14, S17, S18, S19, S20, S21, S24, S25, S26, S29, S30, S32, S34, S35, S44, S45, S46, S48, S49, S53, S54, S55} \hl{fall into} the classification approach.

\textbf{Ranking approach.} The ranking approach learns an ML-based AWI sorter from historical warnings and uses this sorter to prioritize targeted warnings. 
Instead of a binary value in the classification approach, this sorter outputs a continuous value for each targeted warning. 
This continuous value denotes the probability that a targeted warning is actionable, and warnings with higher probabilities are ordered up in the list and can be inspected by developers earlier. 
Thus, the ranking approach is generally considered a regression problem \cite{kiefer1959optimum}.
As no warnings are pruned, the ranking approach does not cause false negatives. 
However, the number of warnings is not reduced and all reported warnings are still required manual inspection. 
In the primary studies, 17 studies \cite{S1, S5, S11, S15, S16, S27, S33, S36, S37, S38, S39, S40, S41, S42, S50, S51, S52} fall into the ranking approach.

\textbf{Combination approach.} The combination approach, involving three studies \cite{S8, S23, S28}, combines classification and ranking approaches to identify targeted warnings. 
Specifically, Yoon et al. \cite{S8} first use the classification approach to classify warnings into actionable and unactionable ones, then prune unactionable warnings, and finally use the ranking approach to prioritize the classified actionable warnings. 
It indicates that this study inherits the weaknesses of the classification approach (i.e., causing false negatives) and the strengths of the ranking approach (i.e., helping developers inspect earlier warnings that are more likely to be actionable).
Two studies \cite{S23, S28} give binary and continuous values to classify and prioritize targeted warnings. 
Due to simultaneously displaying the warning label and warning probability, the two studies can provide more auxiliary information to help developers inspect warnings in comparison to the classification or ranking approach. 
Due to no warning pruning, the two studies do not result in false negatives, while all reported warnings are still manually inspected. 

\begin{tcolorbox}[colback=gray!13, colframe=black, boxrule=0.3mm, boxsep= -0.1cm, middle=-0.1cm]
  \textbf{Summary RQ1}: 
  Based on the warning output format, ML-based AWI approaches can be divided into classification, ranking, and combination approaches. 
  Particularly, the classification approach, covering nearly 61\% of primary studies, is the most commonly used.
\end{tcolorbox}

\subsection{RQ2: What are the characteristics of warning datasets used by ML-based AWI approaches?} \label{rq2}
The warning dataset is a basic component in the ML-based AWI approach.
Based on the typical ML-based AWI workflow in Fig. \ref{fig:workflow}, the warning dataset preparation mainly involves the warning dataset acquisition and labeling.  
To reveal the warning dataset characteristics, we analyze warning dataset acquisition and labeling techniques with associated strengths, weaknesses, and distribution across three categories of ML-based AWI approaches, respectively.

\subsubsection{Warning dataset acquisition.} The warning dataset acquisition is to collect warnings by using SCAs to automatically scan the source code of a project under test.
Thus, the warning dataset acquisition involves the SCA, the development language of the project, and the project source.

\textbf{SCA.} As shown in Table \ref{tab:rq2-sat}, the most commonly used SCA is FindBugs, followed by CppCheck and Sparrow. In all, there are 27 known SCAs mentioned in the primary studies. 
In particular, 12 studies (e.g., \cite{S6, S23}) do not disclose the names of SCAs because these SCAs are only commercially available. 
Also, some studies (e.g., \cite{S27, S33}) use multiple SCAs for the warning dataset acquisition. 
SCAs detect software defects by adopting various techniques.
For example, FindBugs relies on the pattern matching \cite{pattern} to identify potentially dangerous source code. CppCheck uses the flow-sensitive analysis \cite{flow} to reveal undefined program behaviors. Sparrow is designed to detect defects by using the abstract interpretation technique to approximate program behaviors \cite{abstract}. 
SCAs with various techniques could yield different results, especially for the warning category and \hl{amount}. 
As such, the warning dataset acquisition is greatly dependent on a specific SCA.

\textbf{Development language of the project.} Table \ref{tab:rq2-language} shows that most projects focus on Java and C/C++, which account for 49\% (25/51) and 45\% (23/51) of primary studies respectively. Also, a few projects, involving a total of four studies, are PHP, JavaScript, and Solidity. 
In particular, the study \cite{S24} simultaneously pays attention to Java and C/C++ projects. 
Combined with Table \ref{tab:rq2-sat} and Table \ref{tab:rq2-language}, it is observed that the development languages of projects are closely related to SCAs.
It indicates that different development languages of projects are related to the selection of SCAs, thereby affecting the warning dataset acquisition.

\begin{table}[]
    \centering
    \caption{SCA details, including a SCA name with the correspondingly embedded link/reference, commonly supported languages, the fact whether the project under test is required to be compilable before the usage of a SCA (Comp.), and mapping studies of a SCA.}
    \label{tab:rq2-sat}
     \scalebox{0.77}{
    \setlength{\tabcolsep}{0.1mm}{
    \begin{tabular}{|c|c|p{55pt}<{\centering}|c|p{70pt}<{\centering}|| c|c|p{55pt}<{\centering}|c|p{70pt}<{\centering}|}
    \hline
       \textbf{No.} &\textbf{Name}  & \textbf{Languages} & \textbf{Comp.} & \textbf{Studies} & \textbf{No.} &\textbf{Name}  & \textbf{Languages} & \textbf{Comp.} & \textbf{Studies}\\ \hline \hline
1 & \href{https://findbugs.sourceforge.net/}{FindBugs} (16) & Java & Yes & \cite{S1, S2, S3, S4, S9, S10, S18, S19, S20, S21, S28, S32, S33, S44, S48, S55} & 15 &\href{https://pmd.github.io/}{PMD} (1) & Java & No & \cite{S33} \\ \hline
2 &\href{http://cppcheck.sourceforge.net/}{CppCheck} (5)	& C/C++ & No & \cite{S13, S15, S16, S27, S42} & 16 &\href{http://artho.com/jlint/}{Jlint} (1)  & Java & Yes & \cite{S33} \\ \hline
3 &\href{https://github.com/ropas/sparrow}{Sparrow} (4)	& Java, C/C++ & No & \cite{S8, S38, S39, S40} &17 &\href{http://www.jutils.com/}{Lint4J} (1)  & Java & Yes & \cite{S33} \\ \hline
4 & \href{https://www.slideserve.com/eilis/chord-a-program-analysis-platform-for-java}{Chord} (3)	& Java & Yes & \cite{S36, S37, S52} &18 &DTS\cite{DTS} (1)  & C/C++ & Yes & \cite{S46} \\ \hline
5 &\href{https://find-sec-bugs.github.io/}{FindSecBugs} (3) & Java & Yes & \cite{S24, S25, S26} &19 &\href{https://www.cprover.org/cbmc/}{CBMC} (1) & C/C++ & No & \cite{S24} \\ \hline
6 &\href{https://dwheeler.com/flawfinder/}{Flawfinders} (2)	& C/C++ & No & \cite{S15, S16}&20 &\href{https://analysis-tools.dev/tool/jbmc}{JBMC} (1)  & Java & Yes & \cite{S24} \\ \hline
7 &\href{https://code.google.com/archive/p/rough-auditing-tool-for-security/}{RATS} (2) & C/C++, PHP, Python, Perl & No & \cite{S15, S16} &21 &\href{https://b.d4t.cn/RCBJpT}{Rosecheckers} (1)  &C/C++ & No & \cite{S13} \\ \hline
8 &\href{https://analysis-tools.dev/tool/wap}{WAP} (2)	& PHP & No & \cite{S5, S35} &22 &\href{https://www.sonarqube.org/}{SonarQube} (1)  & Java & No & \cite{S12} \\ \hline
9 &\href{https://github.com/oliverklee/pixy}{Pixy} (2)&  PHP & No  & \cite{S5, S35} &23 &\href{https://github.com/manohar9999/SCATE}{SCATE} (1)  & Java, C/C++ & No & \cite{S51} \\ \hline
10 &\href{https://clang-analyzer.llvm.org/}{Clang} (2) & C/C++ & No  & \cite{S27, S42} &24 &\href{http://rips-scanner.sourceforge.net/}{RIPS} (1)  & PHP & No & \cite{S5} \\ \hline
11 &\href{https://frama-c.com/}{Frama-C} (2) & C & No & \cite{S27, S42} &25 &\href{https://github.com/JoseCarlosFonseca/phpSAFE}{phpSAFE} (1)  & PHP & No & \cite{S5} \\ \hline 
12 &\href{https://fbinfer.com/}{Infer} (2) & Java, C/C++, Object-C & No & \cite{S44, S48}&26 &\href{https://d3s.mff.cuni.cz/software/weverca/}{WeVerca} (1)  & PHP & No & \cite{S5} \\ \hline
13 &MC\cite{MC}  (2) & C/C++ &  No &  \cite{S41, S50} &27 & phpMiner\cite{phpminer} (1) & PHP & No & \cite{S35} \\ \hline
14 &\href{http://ropas.snu.ac.kr/2005/airac5/}{Airac} (2)  & C/C++ & No & \cite{S51, S54} &28 &Anonymous (12) & Java, C/C++, Solidity, JavaScript & N/A & \cite{S6, S7, S11, S13, S17, S23, S29, S30, S34, S45, S49, S53} 
\\ \hline
    \end{tabular}
    }
    }
\end{table}

\begin{table}[]
    \centering
    \caption{Development languages of projects under test with corresponding studies.}
    \label{tab:rq2-language}
       \scalebox{0.8}{
    \setlength{\tabcolsep}{0.5mm}{
    \begin{tabular}{|c|p{390pt}<{\centering}|}
    \hline
        \textbf{Language} & \textbf{Studies} \\ \hline \hline
         Java (25) & \cite{S1, S2, S3, S4, S8, S9, S10, S12, S18, S19, S20, S21, S23, S24, S25, S26, S28, S32, S33, S36, S37, S44, S52, S53, S55} \\ \hline
         C/C++ (23) & \cite{S6, S7, S13, S14, S15, S16, S17, S24, S27, S29, S30, S38, S39, S40, S41, S42, S45, S46, S48, S49, S50, S51, S54} \\ \hline
         PHP (2)& \cite{S5, S35} \\ \hline
         JavaScript (1) & \cite{S34}  \\ \hline
         Solidity (1) & \cite{S11} \\ \hline
    \end{tabular}
    }
    }
\end{table}

\textbf{Project source.} By analyzing the primary studies, the project source contains real-world and synthetic categories. 
The project source determines the category of the acquired warning dataset. 
Thus, the warning dataset can be divided into real-world and synthetic categories.

The real-world source refers to the warning dataset collected from the real-world project. 
In the primary studies, the warning datasets in the majority of primary studies (82\%) (i.e., \cite{S2, S3, S4, S6, S7, S9, S10, S12, S17, S18, S19, S20, S21, S29, S30, S32, S34, S35, S44, S45, S46, S48, S1, S5, S11, S15, S16, S33, S36, S37, S38, S39, S40, S41, S50, S51, S52, S8, S23, S28, S54, S55}) are from the real-world source. 
Further, as shown in Fig. \ref{fig:rq2-source}, 24, 15, \hl{3} studies in the real-world source fall into classification, ranking, and combination approaches, respectively. 
In particular, it is observed that the warning dataset, collected from 12 open-source Java projects in the study \cite{S10}, is the most widely used to support ML-based AWI.
Currently, nine studies \cite{S1, S3, S4, S9, S18, S19, S20, S21, S55} have adopted the warning dataset from the study \cite{S10}. 
In the real-world source, the warning dataset can reflect the realistic distribution, while facing the imbalance problem \cite{S4} and missing warnings with uncommon categories.

The synthetic source refers that the warning dataset is collected from the artificially designed projects, including Juliet\footnote{https://samate.nist.gov/SARD/test-suites/112} and OWASP\footnote{https://owasp.org/www-project-benchmark/}. 
Juliet and OWASP focus on C/C++ and Java projects, respectively. 
As shown in Fig. \ref{fig:rq2-source}, seven studies use the warning datasets from the synthetic source. Of these, five studies (i.e., \cite{S13, S14, S26, S49, S53}) fall into the classification approach, and two studies (i.e., \cite{S27, S42}) fall into the ranking approach. 
In comparison to the warning dataset in the real-world source, the warning dataset in the synthetic source can easily remain balanced and cover warnings with uncommon categories, while not depicting the realistic warning distribution.

As shown in Fig. \ref{fig:rq2-source}, two studies \cite{S24, S25} separately perform the warning classification on two categories of warning datasets, where one warning dataset is from the synthetic source (i.e., OWASP) and the other warning dataset is from the real-world source. 
It indicates that the warning datasets in the two studies can inherit the strengths of real-world and synthetic sources.

\subsubsection{Warning dataset labeling.} The warning dataset labeling is to assign labels for warnings. In the primary studies, the warning labeling involves manual, automatic, and hybrid strategies. 

\begin{figure}
    \centering
    \subfloat[Project sources.]{
    \label{fig:rq2-source}
    \includegraphics[scale=0.53]{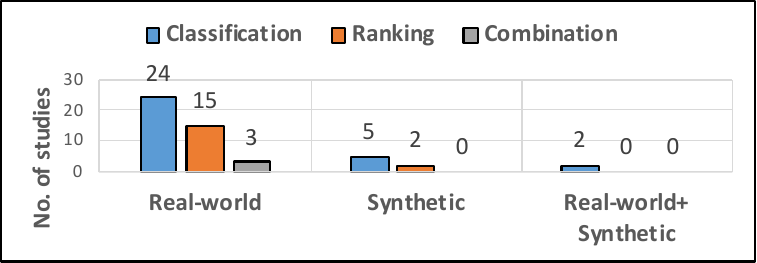}}
    \subfloat[Warning dataset labeling strategies.]{
    \label{fig:rq2-warninglabel}
    \includegraphics[scale=0.53]{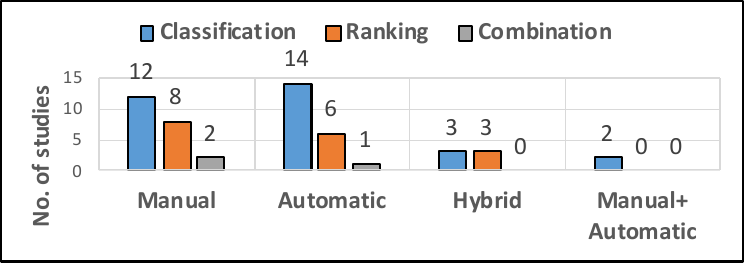}}
    \caption{Distribution of the warning dataset preparation under three categories of ML-based AWI approaches.}
    \label{fig:rq2}
\end{figure}

\textbf{Manual strategy.} The manual strategy \hl{relies on developers' domain knowledge} to label warnings. 
As shown in Fig. \ref{fig:rq2-warninglabel}, the manual strategy involves 22 studies, where 12 \cite{S6, S7, S17, S29, S30, S32, S34, S35, S44, S45, S46, S54}, \hl{8} \cite{S37, S38, S40, S41, S50, S51, S52}, and \hl{2} \cite{S8, S23} studies fall into classification, ranking, and combination approaches respectively. 
The manual strategy can obtain a small number of valuable warnings.
However, the manual strategy has intrinsic limitations.
On the one hand, as illustrated in a study \cite{S22}, it takes an experienced developer five minutes to inspect each warning on average.
It indicates that the manual warning labeling strategy is very time-consuming. 
Consequently, it is difficult for the manual strategy to quickly gather massive warnings.
On the other hand, developers have different levels of experience \cite{myown}. Given the same warning inspected by different developers, the label of this warning may be inconsistent. 
That is, it is error-prone that a warning is manually inspected by only a developer \cite{S9}.

\textbf{Automatic strategy.} The automatic strategy performs the warning labeling without any manual intervention. 
Fig. \ref{fig:rq2-warninglabel} presents that most studies adopt the automatic strategy for warning labeling. 
\hl{Specifically}, 14 \cite{S2, S3, S4, S10, S12, S13, S14, S20, S21, S26, S48, S49, S53, S55}, \hl{6} \cite{S1, S11, S27, S33, S36, S42}, and \hl{1} \cite{S28} studies fall into classification, ranking, and combination approaches respectively. 
Based on different core techniques, the automatic strategy can be further refined into four sub-categories.

First, the warning labels are assigned by judging whether a warning hits the defect-inducing source code. This sub-category is the most commonly used to label warnings in the synthetic source \cite{S13, S14, S26, S27, S42, S49, S53}, because the source code in the synthetic warning dataset has the oracle. That is, if a warning reported by the SCA hits the defect-inducing source code, this warning is labeled to be actionable. Otherwise, this warning is labeled to be unactionable. 
In particular, Liang et al. \cite{S33} use this sub-category to label warnings in the real-world source. 
However, there is no well-prepared defect-inducing source code in the real-world source. To address this problem, Liang et al. increase a preliminary step.
Specifically, it is observed that SCAs tend to detect generic defects rather than project-specific defects. Based on such an observation, the file modification times are used to select generic defect-inducting revisions.
Subsequently, the diff-based algorithm \cite{diff1} is used to identify the defect-inducing source code from these generic revisions. 
It is noted that different studies use different granularities (i.e., statement, method, or file levels) to determine whether a warning hits the defect-inducing source code. 
For example, Alikhashashneh et al. \cite{S14} perform the warning labeling at the method level. That is, a warning is determined to be actionable once this warning falls into the method containing the detect-inducing source code.
More strictly, in the work of Liang et al. \cite{S33}, a warning is considered to be actionable only if (1) the source code lines of this warning hit at least one defect-inducing source code line and (2) this warning in the current revision disappears in a later revision.
The granularity from file to statement levels generally becomes more fine-grained, and the obtained warning labels are more reliable.

Second, the warning labels are obtained by the closed warning-based heuristic, which involves 11 studies \cite{S1, S2, S3, S4, S10, S12, S20, S21, S28, S48, S55}. 
Inspired by the fact that developers are constantly fixing defects in the program, such a heuristic is to give warning labels by performing the warning matching among revisions \cite{S10}. 
Specifically, to judge whether two given warnings are identical in different revisions, the warning matching is performed by comparing the warning characteristics (e.g., category and location).
Then, there are three cases: (1) if a warning in the current revision disappears in any later revision, this warning is considered to be fixed by developers and is labeled to be actionable; (2) if a warning in the current revision is present until the \hl{latest} revision, this warning is considered to be ignored by developers and is labeled to be unactionable; and (3) if the class/method, where a warning is in the current revision, is deleted in a certain later revision, this warning is labeled to be unknown. 
However, the current warning matching is severely affected by the warning-irrelevant source code changes (e.g., the class/method renaming or the code refactoring), thereby making the heuristic \hl{produce} many mislabeled warnings \cite{miningha, ViolationTracker, benchsource}.

Third, the warning labels are obtained by the voting mechanism. Inspired by an observation that a defect is identified exclusively by a single SCA \cite{heihei1, heihei2}, Tran et al. \cite{S11} perform the warning labeling based on the results reported by multiple SCAs.
Specifically, if a warning is only reported by one SCA, this warning is labeled to be unactionable. If a warning is reported by at least two SCAs, this warning is labeled to be actionable. 
\hl{The reliability of a warning being labeled as actionable/unactionable one could be increased due to aggregating the defect detection capabilities of multiple SCAs.
However,} the voting mechanism-based warning labeling strategy could bring mislabeled warnings. 
On the one hand, due to the similar or incorrect warning patterns adopted by SCAs, a warning simultaneously reported by multiple SCAs could be unactionable. On the other hand, due to the complementarity \hl{of} techniques adopted by SCAs \cite{likaixuan}, a warning that is reported by only one SCA may be actionable. 
In addition, compared to the number of warnings reported by only one SCA, the voting mechanism-based warning labeling strategy could report more unactionable warnings because warnings are merged from multiple SCAs \cite{7884656}.

Fourth, the warning labels are obtained via the \emph{k}-object-sensitive version \cite{hgighi}, which mainly consists of the object sensitivity and parameterization framework.
The object sensitivity is a form of context sensitivity for flow-insensitive points-to analysis.
The core idea behind object sensitivity is to separately analyze a method for each object name, which represents a runtime object that may invoke this method.
The parameterization framework, controlled by \emph{k}, is designed for the tradeoff between cost and precision in the object sensitivity. 
For example, Mangal et al. \cite{S36} adopt the k-object-sensitive version (\emph{k} = 4) to assign labels for warnings reported by a context-sensitive but object-insensitive SCA called Chord. 
By leveraging the technique complementarity between Chord and the k-object-sensitive version, the warning labels can be easily obtained. 
However, due to comprising approximations (e.g., flow-insensitivity) in the \emph{k}-object-sensitive version, these warning labels do not have the absolute ground truth.


\textbf{Hybrid strategy.} The hybrid strategy combines manual and automatic strategies to perform warning labeling. 
As shown in Fig. \ref{fig:rq2-warninglabel}, the hybrid strategy involves six studies \cite{S5, S9, S15, S16, S18, S19}, which fall evenly into classification and ranking approaches.
Specifically, two studies \cite{S15, S16} manually inspect whether a method is vulnerable and automatically label warnings by judging whether warnings hit vulnerable methods. 
Yet, the obtained warning labels could be noisy because a vulnerability could be caused by the interprocedural method \cite{S15, S16}.
Similarly, Pereira et al. \cite{S5} adopt the warning dataset in the work of Nunes et al. \cite{dataset2}, which automatically extract the Proof-of-Concept vulnerabilities from CVE\footnote{https://cve.mitre.org/} and manually label warnings by comparing the source code lines of warnings and the vulnerable source code lines. 
However, the source code lines between warnings and vulnerabilities do not correspond perfectly, \hl{and} there could be mislabeled warnings \cite{S5}.
In addition, to alleviate mislabeled warnings caused by the closed warning-based heuristic, Kang et al. \cite{S9} rely on the 2-pass manual inspection to further ensure the reliability of warning labels. 
Subsequently, the warning dataset collected by Kang et al. \cite{S9} is also adopted by other studies \cite{S18, S19}.
The warning labels in Kang et al. \cite{S9} can be considered to be ground-truth. However, these ground-truth warnings only occupy a small part of warnings reported by the SCA because the 2-pass manual inspection resource is limited. 
In summary, the hybrid strategy can to some extent mitigate the weaknesses of manual and automatic strategies. However, it is still a tough task how to combine manual and automatic strategies to improve the warning label quality.

Fig. \ref{fig:rq2-warninglabel} shows that two studies \cite{S24, S25} separately use manual and automatic strategies to label warnings.
Different from the hybrid strategy that simultaneously uses manual and automatic strategies to label warnings on the same warning dataset, the two studies use manual and automatic strategies to label warnings on different warning datasets, respectively. 
Specifically, the manual strategy is used to perform the warning labeling in the real-world source, and the automatic strategy is used to perform the warning labeling in the synthetic source.

\begin{tcolorbox}[colback=gray!13, colframe=black, boxrule=0.3mm, boxsep= -0.1cm, middle=-0.1cm]
  \textbf{Summary RQ2}: The warning dataset preparation stage in the ML-based AWI approaches contains warning dataset acquisition and labeling. 
  In the warning dataset acquisition, it is observed that (1) FindBugs, the most commonly adopted SCA, is used as a research target in 31\% of primary studies; (2) Java and C/C++ are the primarily focused development languages of projects; and (3) the warning dataset from the real-world source occupies 82\% of primary studies. 
  In the warning dataset labeling, the manual and automatic strategies are the most commonly adopted to label warnings, which cover 43\% and 41\% of primary studies, respectively.
\end{tcolorbox}

\subsection{RQ3: What constitutes the warning dataset preprocessing stage of ML-based AWI approaches?} \label{rq3}
The warning dataset preprocessing produces the core input for ML-based AWI. 
Based on the typical ML-based AWI workflow in Fig. \ref{fig:workflow}, 
the warning dataset preparation mainly involves warning feature extraction, warning feature selection, and warning dataset rebalancing.
To illuminate the warning dataset preprocessing, we analyze the techniques used in the three parts, along with strengths, weaknesses, and distribution across three categories of ML-based AWI approaches.

\subsubsection{Warning feature extraction.} This section discusses the warning feature extraction from three perspectives, including extraction sources, extraction ways, and categories of warning features.

\textbf{Warning feature extraction source.} Inspired by the work of Wang et al. \cite{standard}, the warning feature extraction in primary studies can be divided into external, internal, and hybrid sources.

\begin{figure}
    \centering
    \subfloat[Warning feature extraction sources.]{
    \label{fig:rq3-wfsource} \includegraphics[scale=0.53]{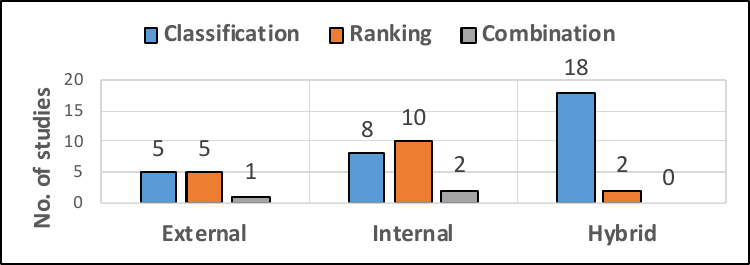}}
    \subfloat[Warning feature extraction ways.]{
    \label{fig:rq3-wfeway}
\includegraphics[scale=0.53]{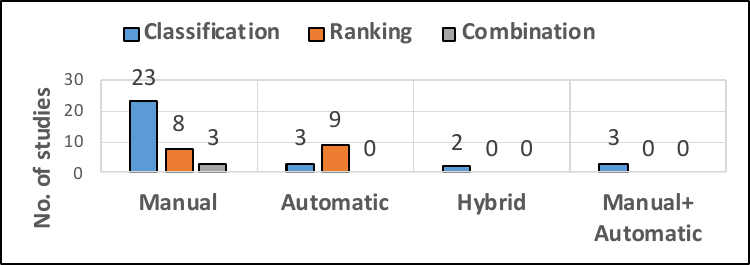}}
    \hfill
    \subfloat[Warning feature categories. Cha, Sta, Con, Seq, and Str are characteristic-based, statistical, content-based, sequential, and structural categories, respectively. Studies that compare the AWI performance differences of multiple warning feature categories are merged together (e.g., Cha+Sta).]{
    \label{fig:rq3-wfcategory}
    \includegraphics[scale=0.53]{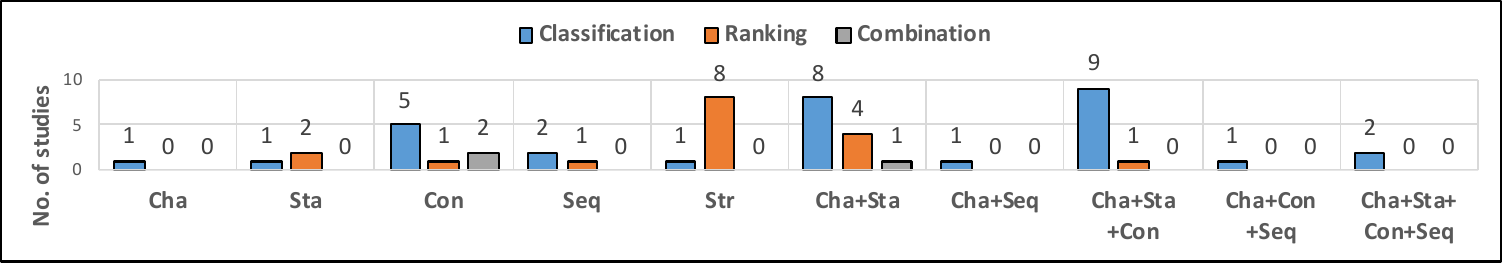}}
    \caption{Distribution of the warning feature extraction under three categories of ML-based AWI approaches.}
    \label{fig:rq3-wfe}
\end{figure}

The external source is to extract warning features from the SCA report or developer feedback. 
Fig. \ref{fig:rq3-wfsource} shows that the external source has 11 studies, where \hl{5} \cite{S6, S17, S34, S48, S49}, \hl{5} \cite{S5, S27, S41, S42, S50}, and \hl{1} \cite{S23} studies involve classification, ranking, and combination approaches respectively. 
The warning reported by SCAs contains a series of warning characteristics (e.g., category, severity, and location). 
Many studies (e.g., \cite{S34}) extract the original warning characteristics from the SCA report as warning features. 
Also, some studies (e.g., \cite{S5, S17}) derive some new warning features (e.g., the number of warnings in a class/method) from the original warning characteristics. 
In addition, a few studies \cite{S6, S10} extract warning features from the developer feedback on warnings reported by the SCA.
The warning features in the external source are obtained by only parsing the SCA report or the developer feedback. Thus, the warning features are generally easily extracted. 
However, due to independent of the warning-related source code, the warning features in the external source only capture the shallow warning information and fail to reveal the root cause of the warning.

The internal source is to extract warning features from the source code containing warnings. 
Fig. \ref{fig:rq3-wfsource} shows that the internal source occupies 39\% (20/51) of primary studies, where \hl{8} \cite{S7, S14, S26, S30, S35, S45, S53, S54}, 10 \cite{S11, S15, S16, S36, S37, S38, S39, S40, S51, S52}, and \hl{2} \cite{S8, S28} studies fall into classification, ranking, and combination approaches respectively.
Given a warning reported by the SCA, the warning location (i.e., the class and method information containing this warning and warning line numbers) are generally displayed. 
Based on the warning location, the source code related to the class, method, and warning line numbers can be obtained. 
Besides, by using program analysis techniques (e.g., program slicing \cite{weiser1984program}), the source code that traces the path leading to the warning is available.
Moreover, based on the software evolution history, the warning-related software change history can be recorded in terms of class, method, and warning lines.
For example, Lee et al. \cite{S7} extract features from the source code surrounding warning line numbers. 
Some studies \cite{S15, S16} use the abstract syntax tree with control flow construction technique to extract features from the source code containing warnings. 
Compared to warning features in the external source, the warning features in the internal source could capture deeper information. 
However, there are some limitations in the internal warning feature extraction source. 
On the one hand, the warning features in the internal source are more difficult to be extracted than those of the external source. 
On the other hand, the source code indeed contains rich syntactic and structural information of the warning, but not all of the source code in the project is related to the warning. Thus, to capture comprehensive warning information while eliminating warning-irrelevant information, it is tough how to appropriately extract warning features from the internal source.

The hybrid source is to extract warning features from external and internal sources. 
Fig. \ref{fig:rq3-wfsource} presents 20 studies in the hybrid source, where 18 \cite{S2, S3, S4, S9, S10, S12, S13, S18, S19, S20, S21, S24, S25, S29, S32, S44, S46, S55} and \hl{2} \cite{S1, S33} studies belong to classification and ranking approaches respectively.
For example, Hegedűs et al. \cite{S12} combine the warning category extracted from the SCA report and the source code \hl{surrounding} warning line numbers as warning features. 
Wang et al. \cite{S10} extract eight categories of warning features (e.g., warning characteristics and software change history) from the SCA report and source code containing the warning. 
The hybrid source can represent warnings comprehensively \hl{by} combining warning features with external and internal sources. 
However, the existing studies directly concatenate warning features from external and internal sources together. 
Actually, there is a redundancy in the warning features in external and internal sources \cite{S18}, which severely hinders the AWI performance.
Thus, it is investigated how to reasonably combine warning features from external and internal sources.

\textbf{Warning feature extraction way.} Based on whether there is manual intervention \cite{follow}, the warning feature extraction involves manual, automatic, and hybrid ways.

The manual way is to rely on experts' domain knowledge for warning feature extraction. Fig. \ref{fig:rq3-wfeway} shows that such a way involves the most studies, where 23 \cite{S2, S3, S4, S6, S9, S10, S13, S14, S17, S18, S19, S20, S21, S29, S30, S32, S34, S35, S46, S48, S49, S54, S55}, \hl{8} \cite{S1, S5, S27, S33, S41, S42, S50, S51}, and \hl{3} \cite{S8, S23, S28} studies fall into classification, ranking, and combination approaches respectively. 
By the manual way, a small number of meaningful features can be obtained and be suitable for homogeneous warnings.
For example, Zheng et al. \cite{S48} assume that the complex source code (e.g., more condition statements and lines of code) is more likely to be found bugs by the SCA.  
Based on this assumption, they manually extract 25 features (e.g., frequency of OR/AND conditions in the warning trace and length of warning line numbers) to denote the complexity of \hl{the warning-related} source code for AWI.
However, it is tedious to manually extract warning features. Besides, different experts could yield heterogeneous domain knowledge, thereby causing biased warning features. 
More fatally, the manually extracted warning features become obsolete over time, which makes it difficult to identify newly reported warnings due to the concept drift \cite{gama2014survey}.

The automatic way is to leverage Deep Learning (DL) or datalog reasoning techniques for warning feature extraction. Fig. \ref{fig:rq3-wfeway} shows 12 studies in the automatic way, where \hl{3} \cite{S7, S26, S53} and \hl{9} \cite{S11, S15, S16, S36, S37, S38, S39, S40, S52} studies cover classification and ranking approaches respectively. 
Specifically, six studies \cite{S7, S11, S15, S16, S26, S53} represent warnings by using DL techniques (e.g., Word2Vec) to encode the warning-related source code.
The remaining six studies \cite{S36, S37, S38, S39, S40, S52} extract the datalog derivation graph from the warning-related source code via def-use analysis \cite{defuse} and perform queries for warnings by applying the Bayesian inference algorithm to such a graph. 
The DL technique can automatically capture the deep warning information. However, the warning features obtained by DL could be sparse and high-dimension as well as require a lot of computing power and massive historical warnings.
By contrast, despite maintaining the declarative semantics and logical consistency based on the flexible rule definition in the def-use analysis, the datalog reasoning technique suffers from limited expressiveness in the program with the complex data types and the high computational overhead \cite{motik2006comparison}.


The hybrid way is to extract features by combining the manual and automatic ways. Fig. \ref{fig:rq3-wfeway} signifies that two studies \cite{S12, S45} related to the classification approach use the hybrid way for warning feature extraction.
Specifically, Hegedűs et al. \cite{S12} manually extract warning characteristics and automatically extract the source code surrounding the warning line numbers, thereby concatenating warning characteristics and source code as warning features.  
Meng et al. \cite{S45} first automatically construct the code property graph from the \hl{warning-related} source code and manually extract warning features from this code property graph. 
Compared to warning features in the manual way, the warning features in the hybrid way could be more robust for AWI due to embracing the automatic extraction of warning features.
Compared to the warning features in the automatic way, the warning features in the hybrid way are \hl{require} additional domain knowledge due to relying on the manual way to extract warning features.

Fig. \ref{fig:rq3-wfeway} signifies that three studies \cite{S24, S25, S44} related to the warning classification separately extract warning features in manual and automatic ways.
For example, Kharkar et al. \cite{S44} manually extract identifiers related to the null dereference and resource leak (e.g., whether \emph{return null} exists) from the local and non-local warning-related source code as warning features for AWI.  
Additionally, they automatically extract the local and non-local warning-related source code as warning features and apply the DL model to encode these warning features for AWI.
Subsequently, they compare the AWI performance difference under warning features between manual and automatic ways. 


\textbf{Warning feature category.} By analyzing the primary studies, the warning features can be classified into characteristic-based, statistical, content-based, sequential, and structural categories. Table \ref{tab:wfcategory} shows the details of different warning feature categories.

The characteristic-based category denotes the warning characteristics extracted from the external source, especially the SCA report. 
The warning features (e.g., category and severity) in the characteristic-based category are commonly used by prior studies (e.g., \cite{S6, S12, S49}).
It is noted that some SCAs (e.g., Infer) report warnings with associated source code. 
As such, the warning features (e.g., the source and sink identifiers) extracted from such the source code \cite{S34, S48} are also considered as the characteristic-based category. 
In addition, a few studies \cite{S6, S10} extract the developer idea (i.e., the developer feedback on warnings reported by the SCA) as the characteristic-based warning feature.
As shown in Fig. \ref{fig:rq3-wfcategory}, the characteristic-based category involves 28 studies, where one study \cite{S49} only uses the characteristic-based warning features for AWI.
The characteristic-based warning features are easily extracted. 
\hl{However, due to missing a deep understanding of the warning-related source code,} the characteristic-based warning features could lack sufficiently discriminative capability when separately used for AWI.

The statistical category represents the warning statistics information from the SCA report or the source code containing warnings. 
The warning features from the SCA report (e.g., the number of warnings in a class/method) denote the warning aggregation information \cite{S2, S10}. 
The warning features from the source code containing warnings (e.g., the depth and modification times of a method containing warnings) denote the warning complexity and evolution information \cite{S33}. 
As shown in Fig. \ref{fig:rq3-wfcategory}, the statistical category involves 28 studies, where three studies \cite{S14, S41, S50} separately adopt the statistical warning features for AWI.
The statistical warning features unveil the hidden mathematical distribution of warnings. 
When there are sufficient history warnings, it is very instructive to identify newly reported warnings \cite{guide1, guide2, guide3, S48}. 
However, the extraction process of the statistical warning features requires intensive computation power. 
\hl{Also, it is worthy noting that due to the incorrect or inappropriate implementation ways,} some statistical warning features (e.g., the defect density for warning type) yield data leakage \cite{S9}, which could cause exaggeration of the ML-based AWI performance.

\begin{table}[]
    \centering
    \caption{Warning feature categories with their associated descriptions, strengths, and weaknesses.}
    \label{tab:wfcategory}
           \scalebox{0.6}{
    \setlength{\tabcolsep}{0.1mm}{
    \begin{tabular}{|p{40pt}<{\centering}|p{320pt}<{\centering}|p{150pt}<{\centering}|p{140pt}<{\centering}|}
    \hline
        \textbf{Category}	&	\textbf{Description}	&	\textbf{Strengths}	&	\textbf{Weaknesses}	\\	\hline \hline
\makecell{Charac-\\teristic}	&	This category denotes the warning characteristics extracted from the external source, i.e., the SCA report and developer feedback on the reported warnings.	&	Being easily extracted	&	
\hl{Lacking sufficiently discriminative capability due to missing a deep understanding of the warning-related source code.}
\\	\hline
Statistical	& This category denotes the warning statistics information from two aspects. Features (e.g., the number of warnings in a class) from the SCA report provide aggregated warning information. Features (e.g., the depth and modification times of a class) from the source code containing warnings provide complexity and evolution information.	&	 Unveiling the hidden mathematical distribution of warnings, which is instructive to identify warnings when there are sufficient history warnings	&	Requiring intensive computation power; \hl{Being likely to having} the data leak in some warning features \hl{due to the implementation mistakes}	\\	\hline
Content	&	This category signifies important identifiers in the source code containing warnings,  e.g., identifiers related to loops. Unlike the characteristic-based category that focuses on the SCA report, the content-based category focuses on the source code containing warnings.	&	Capturing warnings with regular patterns due to searching important identifiers from the source code containing warnings	&	Heavily relying on domain knowledge and being labor-intensive	\\	\hline
\makecell{Sequen-\\tial}	&	This category denotes the sequential information of the source code containing warnings by considering the source code containing warnings as the natural language text and using DL techniques to encode such the source code. Unlike the content-based category that \hl{extracts} the important identifiers from the source code containing warnings, the sequential category aims to learn the sequential distribution from the source code containing warnings.	&	Capturing the sequential information of warnings	&	Missing the structural information of warnings; Being generally sparse and high-dimension; Being difficult to interpret	\\	\hline
Structural	&	This category embodies the structural information of warnings by mainly extracting warning features from the abstract syntax tree, control flow, or data flow of warnings. Unlike the sequential category that considers the source code containing warnings as the natural language text, the structural category considers the source code containing warnings as the tree/graph structure. 	&	Capturing richer information with the warning syntax and semantics	&	Being difficult to interpret when DL techniques are used for AWI; Enduring limited expressiveness when the datalog derivation graph is used for AWI	\\	\hline
    \end{tabular}
    }
    }
\end{table}

The content-based category signifies important identifiers in the source code containing warnings. 
The main process to extract the content-based warning features is shown as follows. 
First, the source code containing warnings is obtained by establishing the abstract syntax tree \cite{S8, S28, S24, S46}, code property graph \cite{S45}, program dependency graph \cite{S24, S25}, \hl{local/non-local} context \cite{S44}, or loops-related context \cite{S30}.
Second, the important identifiers (e.g., identifiers related to loops and library calls \cite{S30}) are extracted from the obtained source code. 
In general, the first step is an automatic process, and the second step is a manually designed process.  
In particular, two studies \cite{S24, S25} automatically extract the content-based warning features for AWI, i.e., applying the Bag-of-Word model \cite{zhang2010understanding} to automatically calculate the frequency or occurrence of each identifier in the source code containing warnings. 
Noted, the difference between the characteristic-based and content-based categories is that the former mainly focuses on the SCA report, and the latter focuses on the source code containing warnings. 
Fig. \ref{fig:rq3-wfcategory} presents that 21 studies fall into the content-based category. Of these, eight studies only adopt the content-based warning features for AWI, where five \cite{S30, S35, S45, S46, S54}, one \cite{S51}, and two \cite{S8, S28} studies are related to classification, ranking, and combination approaches respectively.
The content-based warning features can capture warnings with regular patterns due to searching important identifiers from the source code containing warnings, while heavily relying on domain knowledge and being labor-intensive.

The sequential category denotes the sequential information of the source code containing warnings.
The sequential warning features are extracted \hl{in} two steps. 
First, the source code containing warnings is obtained by performing the program slicing \cite{S24, S25, S26} or extracting the source code surrounding the warning line numbers \cite{S7, S11, S12}. 
Second, such source code is considered as the natural language text and is encoded into vectors via DL techniques. 
Noted, the difference between the content-based and sequential categories is that the former aims to extract the important identifiers from the source code containing warnings, and the latter aims to learn the sequential distribution from the source code containing warnings.
Fig. \ref{fig:rq3-wfcategory} describes that there are seven studies in the sequential category, where three studies \cite{S7, S11, S26} separately use the sequential warning features for AWI. 
The sequential warning features can capture the sequential information of warnings while missing the structural information of warnings.
Also, the sequential warning features extracted by DL techniques are generally sparse and high-dimension. 
Moreover, it is difficult to interpret the role of these warning features in AWI due to the inherent limitations of DL in terms of explainability \cite{interpret}.

The structural category embodies the structural information of warnings by mainly extracting warning features from the abstract syntax tree, control flow, or data flow of warnings. 
Fig. \ref{fig:rq3-wfcategory} shows that one study \cite{S53} falls into the classification approach and eight studies \cite{S15, S16, S36, S37, S38, S39, S40, S52} belong to the ranking approach.
Specifically, three studies \cite{S15, S16, S53} use DL techniques to automatically learn the structural distribution information from the program dependency graph and control flow-based abstract syntax tree of warnings. 
Six studies \cite{S36, S37, S38, S39, S40, S52} automatically reason the root cause of warnings by using def-use analysis \cite{defuse} to extract the datalog derivation graph.
Noted, the difference between the sequential and structural categories is that the former considers the source code containing warnings as the natural language text, and the latter considers the source code containing warnings as the tree/graph structure. 
As such, compared to the sequential warning features, the structural warning features can capture richer information with the warning syntax and semantics. 
However, in the studies \cite{S15, S16, S53} that rely on DL techniques to represent warnings for AWI, the structural warning features face the same limitation as those of the sequential category, i.e., difficult to explain \cite{interpret}. 
In the studies \cite{S36, S37, S38, S39, S40, S52} that adopt the datalog reasoning technique to represent warnings for AWI, the structural warning features endure limited expressiveness when encountering the program with the complex data types \cite{motik2006comparison}.

\subsubsection{Warning feature selection.} 
Feature selection is an important but optional process in the ML-based tasks \cite{follow}. 
By analyzing the primary studies, the warning feature selection can be classified into filter, wrapper, embedded, and dimension reduction techniques.
The filter technique independently ranks features by using statistical metrics without involving any ML model.
The wrapper technique measures feature subsets by training and testing a specific ML model and uses its performance as the evaluation metric.
The embedded technique determines the best features by integrating feature selection into the model training process.
Different from the above three techniques that aim to select a subset of the original features, the dimension reduction technique reduces the number of features by transforming the feature space.

\begin{figure}
    \centering
    \subfloat[Warning feature selection techniques.]{
    \label{fig:rq3-wfselection}
    \includegraphics[scale=0.53]{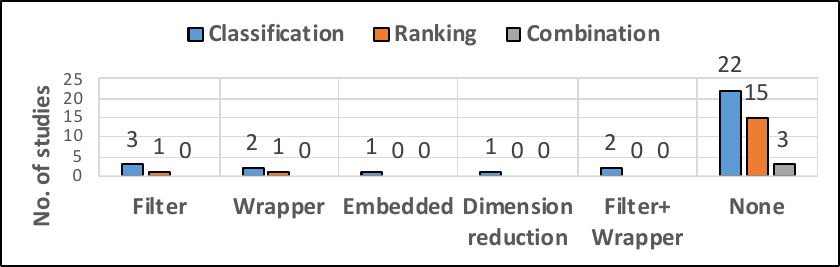}}
    \subfloat[Warning dataset rebalancing techniques.]{
    \label{fig:rq3-wdrebalance}
    \includegraphics[scale=0.53]{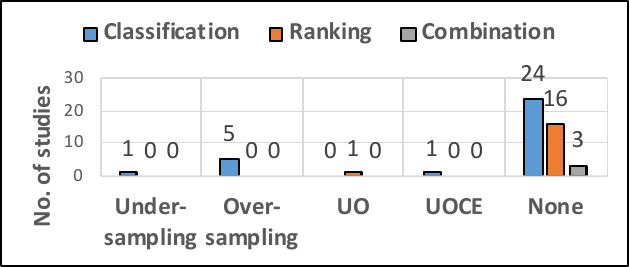}}
    \hfill
    \caption{Distribution of the warning feature selection and warning dataset rebalancing techniques under three categories of \hl{ML-based AWI} approaches. In Fig. \ref{fig:rq3-wdrebalance}, \hl{``UO'' is to use undersampling and oversampling for AWI. ``UOCE'' is to use four sampling techniques for AWI.}
    }
    \label{fig:rq3-wfselectionrebalance}
\end{figure}

\textbf{Filter technique.} The filter technique selects the warning feature subset by calculating the feature importance or correlation via a certain criterion.
Further, based on whether warning labels are required, the warning feature selection involves unsupervised and supervised ways \cite{S18}. 
As shown in Fig. \ref{fig:rq3-wfselection}, four studies \cite{S5, S14, S18, S55} attempt the filter technique for warning feature selection.
For example, Pereira et al. \cite{S5} separately use the correlation- and variance-based feature selection techniques for AWI, where the correlation-based one is supervised and the variance-based one is unsupervised.
In particular, UNEASE \cite{S18}, the first unsupervised warning feature selection technique for AWI, mainly performs the feature clustering and feature ranking to select a warning feature subset from the original warning feature set. 
The filter technique is computationally fast, simple to implement, and model-independent. However, the determination of the warning feature subset in the filter technique often requires manually setting thresholds. Also, due to \hl{model independence}, the filter technique could ignore model biases and cause low AWI performance.

\textbf{Wrapper technique.} The wrapper technique selects the best warning features by evaluating the ML-based AWI model performance.
Fig. \ref{fig:rq3-wfselection} shows the wrapper technique in three studies, where two \cite{S10, S54} and one \cite{S1} studies fall into classification and ranking approaches, respectively. 
For example, Wang et al. \cite{S10} use the greedy backward elimination algorithm for feature selection. Such an algorithm iteratively trains an ML-based AWI classifier, greedily removes features one by one from the original warning features to maximize the performance of this classifier, and finally returns the warning feature subset with the best AWI performance.
The wrapper technique can consider dependencies among warning features \cite{follow}, as it selects the warning feature subset by training and testing an ML-based AWI model, which inherently reflects the interaction among warning features.
Thus, the obtained warning feature subset could achieve good AWI performance. 
However, compared to the filter technique, the wrapper technique is model-dependent and computationally intensive due to the requirement to train an AWI model multiple times.

\textbf{Embedded technique.} The embedded technique automatically performs the warning feature selection when constructing the ML-based AWI model. 
The embedded technique only involves one study \cite{S48} with the classification approach. 
Specifically, to explain the role of 25 extracted features in AWI, Zheng et al. \cite{S48} calculate the warning feature importance via Random Forest (RF). 
Once RF is well-constructed for AWI, the warning feature importance is obtained.
Finally, the root node in RF (\hl{e.g.}, the line number of the error) is dominant for AWI. 
The embedded technique can select optimal features in parallel to training the ML-based AWI model. Thus, the embedded technique mitigates the weaknesses of the filter and wrapper techniques. 
However, the embedded technique is model-specific and requires an understanding of the model details.

\textbf{Dimension reduction technique.} The dimension reduction technique maps the original warning feature space to a low-dimensional space via projection. 
The dimension reduction technique only contains one study \cite{S21}, which leverages the fractal-based method \cite{levina2004maximum} to convert the high-dimension warning features into a more compressed space without the principal information loss. 
Compared to the typical dimension reduction technique (i.e., Principal Component Analysis, PCA), Yang et al. \cite{S21} consider that the fractal-based method is more appropriate for AWI due to (1) better handling the increasingly sophisticated and non-linearly decomposable data and (2) no need for manually setting thresholds.
Besides, the dimension reduction technique is computationally fast because it is model-independent and unsupervised.
However, the warning features determined by the dimension reduction technique are difficult to interpret.

Fig. \ref{fig:rq3-wfselection} shows that 11 studies adopt the warning feature selection techniques for AWI, \hl{of which} two studies \cite{S2, S6} separately use filter and wrapper techniques for warning feature selection.
However, the remaining 40 studies do not adopt any warning feature selection technique for AWI.

\subsubsection{Warning dataset rebalancing.} Class imbalance, where the number of positive (i.e., actionable) samples is much fewer than that of negative (i.e., unactionable) samples, is a common phenomenon in the warning dataset used for the ML-based AWI model training \cite{S4}. 
To mitigate the above phenomenon, some primary studies adopt the warning dataset rebalancing technique for AWI. 
By analyzing these primary studies, the warning dataset rebalancing technique involves undersampling, oversampling, combined sampling, and ensemble sampling. 

\textbf{Undersampling technique.} The undersampling technique mitigates the class imbalance by eliminating unactionable warnings from the original warning dataset. 
Fig. \ref{fig:rq3-wdrebalance} describes that only one study \cite{S20} related to the classification approach adopts the aggressive undersampling to rebalance the warning dataset. 
Different from the random undersampling \cite{S4, S5}, the aggressive undersampling throws away unactionable warnings close to the decision boundary of the ML-based AWI model and accesses actionable warnings until the ratio of actionable and unactionable warnings is balanced in the training set.
The undersampling can easily and quickly obtain the balanced classes. However, the undersampling could miss potentially valuable information due to throwing away the majority of unactionable warnings.

\textbf{Oversampling technique.} The oversampling creates a superset of the original warning dataset by generating actionable warnings. 
As shown in Fig. \ref{fig:rq3-wdrebalance}, five studies \cite{S3, S12, S14, S19, S35} use the oversampling to rebalance the warning dataset when performing the warning classification.
The commonly used oversampling technique is the Synthetic Minority Oversampling technique (SMOTE), which involves four warning classification studies \cite{S3, S14, S19, S35}.
Similar to the undersampling, the oversampling can also easily and quickly rebalance the warning dataset. However, the oversampling increases the likelihood of overfitting in the ML-based AWI model \cite{overfitting}.

\textbf{Combined sampling technique.} The combined sampling is constructed by combining oversampling and undersampling. 
Only \hl{one} study \cite{S4} attempts two commonly used combined sampling techniques (i.e., EditNearestNeighbors and SMOTEENN) to perform the warning rebalancing for AWI.
The combined sampling can mitigate the weaknesses of undersampling and oversampling techniques. 
However, compared to the undersampling and oversampling, the combined sampling is more complex and time-consuming.

\textbf{Ensemble sampling technique.} The ensemble sampling solves the class imbalance by embedding the aforementioned sampling techniques into the ensemble learning models. 
The commonly used ensemble sampling is EasyEnsemble, RUSBoost, BalancedBagging, and BalancedRandomForest \cite{S4}.
In general, ensemble sampling can help achieve a high AWI performance.
However, the ensemble sampling is model-dependent and requires \hl{understanding} the model details. 

As shown in Fig. \ref{fig:rq3-wdrebalance}, two studies \cite{S4, S5} attempt different sampling techniques for the warning dataset rebalancing. 
Specifically, Pereira et al. \cite{S5} \hl{separately} use undersampling and oversampling to mitigate the class imbalance, thereby performing the warning ranking. 
Ge et al. \cite{S4} evaluate the impact of the above four sampling techniques on the ML-based AWI performance. 
Additionally, 43 studies do not adopt any warning dataset rebalancing technique for AWI.

\begin{tcolorbox}[colback=gray!13, colframe=black, boxrule=0.3mm, boxsep= -0.1cm, middle=-0.1cm]
  \textbf{Summary RQ3}: The warning dataset preprocessing stage in the ML-based AWI approaches contains warning feature extraction, warning feature selection, and warning dataset rebalancing. 
  Specifically, it is observed in the warning feature extraction that (1) the main sources are internal and hybrid ones; (2) the manual way is the most prevalent, which occupies 67\% of primary studies; (3) \hl{the} most commonly used warning features fall into the characteristic-based and statistical categories. Particularly, 53\% of primary studies combine multiple categories of warning features for AWI.
  In addition, only 22\% and 16\% of primary studies attempt the warning feature selection and warning dataset rebalancing techniques for AWI, respectively.
\end{tcolorbox}

\subsection{RQ4: What model construction techniques are used by ML-based AWI approaches?} \label{rq4} 
The ML-based AWI model construction is the key step to provide the decision support for a newly reported warning.
By following the typical ML-based AWI workflow in Fig. \ref{fig:workflow}, 
we present the answer to this RQ by enlisting the techniques used in the model category, learning category, AWI model structure, and AWI construction scenario, along with associated strengths, weaknesses, and distribution across three categories of ML-based AWI approaches.

\subsubsection{Model category in AWI}  
The ML categories used for AWI can be divided into Traditional ML (TML), Deep Learning (DL), statistical inference, and Pre-Trained Model (PTM).

\textbf{TML model.} Fig. \ref{fig:rq4-modelcategory} shows that 26 studies only use TML models for AWI. 
Table \ref{tab:rq4-modelcategory} describes that RF (22/51) are the most commonly used ML models, followed by DT and SVM.

\textbf{DL model.} Table \ref{tab:rq4-modelcategory} shows eight DL models used for AWI, where CNN occupies the largest proportion. 
Fig. \ref{fig:rq4-modelcategory} shows that four studies only use DL models for AWI, where one \cite{S7} and three \cite{S11, S15, S16} studies are related to classification and ranking approaches, respectively.

\begin{table}[]
    \centering
    \caption{Distribution of the model category. The full names of model abbreviations are shown \hl{below}: Random Forest (RF), Decision Tree (DT), Naive Bayes (NB), Support Vector Machine (SVM), Logistical Regression (LR), K-Nearest Neighbor (KNN), Bayesian Network (BN), KStar (K*), Ensemble (the customized model based on the idea of ensemble learning), Random Committee (RC), Random Tree (RT), Others (i.e., PART, Ridor, Conjunctive Rule, ADTree, REPTree, LMT, LWL, IBK, and Decision Table), 
    Convolutional Neural Network (CNN), Long Short-Term Memory (LSTM), MultiLayer Perceptron (MLP), Deep Neural Network (DNN), Bi-directional Gating Recurrent Unit (BiGRU), Gated Graph Neural Network (GGNN), Bi-directional Long Short-Term Memory (BiLSTM), Artificial Neural Network (ANN), and Markov Logic Network (MLN).}
    \label{tab:rq4-modelcategory}
    \scalebox{0.85}{
    \setlength{\tabcolsep}{0.1mm}{
    \begin{tabular}{|p{42pt}<{\centering}|p{64pt}<{\centering}|p{35pt}<{\centering}|p{29pt}<{\centering}|p{38pt}<{\centering}|p{250pt}<{\centering}|}
    \hline
      \multirow{2}{*}{\textbf{\makecell{Model \\ category}}} 	&	\multirow{2}{*}{\textbf{Model}}	&	\multicolumn{3}{c|}{\textbf{Approaches}} & \multirow{2}{*}{\textbf{Studies}} \\ \cline{3-5}
     & & \textbf{\makecell{Classi-\\fication}}	&	\textbf{\makecell{Ran-\\king}}	&	\textbf{
    \makecell{Combi-\\nation}}	&	\\ \hline \hline
\multirow{20}{*}{\rotatebox{90}{TML model}}	
&	RF(22)	&	21	&	0	&	1	&	\cite{S12, S21, S23, S24, S25, S53, S3, S4, S6, S10, S13, S14, S18,S17, S20, S46, S29, S35, S48, S49, S54, S55} \\ \cline{2-6}
&	DT(21)	&	15	&	4	&	2	&	\cite{S5, S12, S19, S21, S23, S24, S25, S1, S2, S3, S4, S6, S10, S18, S20, S27, S28, S34, S35, S42, S49} \\ \cline{2-6}
&	SVM(18)	&	15	&	1	&	2	&	\cite{S19, S21, S23, S24, S25, S1, S4, S8, S9, S10, S14, S18, S20, S34, S35, S45, S55} \\ \cline{2-6}
&	NB(17)	&	14	&	1	&	2	&	\cite{S12, S23, S24, S25, S26, S53, S1, S2, S4, S6, S10, S18, S28, S34, S35, S45, S54} \\ \cline{2-6}
&	LR(13)	&	12	&	1	&	0	&	\cite{S19, S44, S1, S2, S4,S10, S13, S18, S29, S32, S35, S49, S54} \\ \cline{2-6}
	&	KNN(9)	&	7	&	2	&	0	&	\cite{S53, S1, S4, S9, S14, S18, S33, S35, S45}	\\ \cline{2-6}
 &	Boosting(9)	&	7	&	2	&	0	&	\cite{S4, S10, S27, S42, S45, S13, S29, S48, S54} \\ \cline{2-6}
	&	BN(6)	&	5	&	0	&	1	&	\cite{S24, S25, S2, S28, S34, S35} 	\\ \cline{2-6}
 &	K*(4)	&	4	&	0	&	0	&	\cite{S24, S25, S2, S34}	\\ \cline{2-6}
	&	OneR(3)	&	3	&	0	&	0	&	\cite{S24, S25, S34}	\\ \cline{2-6}
	&	ZeroR(3)	&	3	&	0	&	0	&	\cite{S24, S25, S34}	\\ \cline{2-6}
 &	Ensemble(3)	&	3	&	0	&	0	&	\cite{S46, S48, S17}	\\ \cline{2-6}
	&	RC(2)	&	2	&	0	&	0	&	\cite{S6, S17}	\\ \cline{2-6}
	&	LightGBM(2)	&	2	&	0	&	0	&	\cite{S13, S48}	\\ \cline{2-6}
 &	NDTree(1)	&	1	&	0	&	0	&	\cite{S34}	\\ \cline{2-6}
	&	Ripper (1)	&	1	&	0	&	0	&	\cite{S14}	\\ \cline{2-6}
	&	DTNB(1)	&	1	&	0	&	0	&	\cite{S17}	\\ \cline{2-6}
 &	RT(1)	&	1	&	0	&	0	&	\cite{S35}	\\ \cline{2-6}
	&	Others(1)	&	1	&	0	&	0	&	\cite{S2}	\\ \hline
\multirow{9}{*}{\rotatebox{90}{DL model}}		
&	CNN(6)	&	4	&	2	&	0	&	\cite{S7, S11, S19, S21, S15, S53}	\\ \cline{2-6} 
	&	LSTM(4)	&	4	&	0	&	0	&	\cite{S24, S25, S26, S53}	\\ \cline{2-6}
	&	MLP(4)	&	3	&	0	&	1	&	\cite{S23, S24, S25, S35}	\\ \cline{2-6}
	&	DNN(2)	&	2	&	0	&	0	&	\cite{S12, S21}	\\ \cline{2-6}
	&	BiGRU(2)	&	0	&	1	&	0	&	\cite{S15}	\\ \cline{2-6}
 &	GGNN(2)	&	2	&	0	&	0	&	\cite{S24, S25} \\ \cline{2-6}
	&	BiLSTM(2)	&	0	&	2	&	0	&	\cite{S11, S16}	\\ \cline{2-6} 
 &	ANN(3)	&	2	&	1	&	0	&	\cite{S5, S19, S55}	\\ \hline
 \multirow{4}{*}{\rotatebox{90}{\makecell{Statistical \\ inference \\model}}}	&	Bayesian inference (7)	&	0	&	7	&	0	&	\cite{S36, S37, S38, S39, S40, S41, S51}	\\ \cline{2-6}
	&	Z-test(1)	&	0	&	1	&	0	&	\cite{S50}	\\ \cline{2-6}
	&	MLN(1)	&	0	&	1	&	0	&	\cite{S52}	\\ \hline
 \multirow{3}{*}{\rotatebox{90}{PTM}} 	&	\makecell{CodeBERTa(1)} 	&	1	&	0	&	0	&	\cite{S44}	\\ \cline{2-6}
 &	\makecell{CodeBERT(1)} 	&	1	&	0	&	0	&	\cite{S19}	\\ \cline{2-6}
 & GPT-C(1)	&	1	&	0	&	0	&	\cite{S44}	\\ \hline
    \end{tabular}
    }
    }
\end{table}

\textbf{Statistical inference model.} Fig. \ref{fig:rq4-modelcategory} shows that the statistical inference model is adopted by nine studies. 
Specifically, six studies \cite{S36, S37, S38, S39, S40, S41} convert the datalog derivation graph of warnings into a \hl{Bayesian} inference model and utilize the real-time feedback of developers to adjust this model for AWI. 
Unlike the six studies, Jung et al. \cite{S51} directly construct a \hl{Bayesian} inference model for AWI without any real-time adjustment.
In addition, Kremenek et al. \cite{S50} rank warnings by applying z-test statistics to the number of historical warnings. Zhang et al. \cite{S52} rely on MLN to convert the datalog derivation graph of warnings into a warning prioritization model for AWI.

\textbf{PTM.} PTM performs the self-supervised training on large-scale and unlabeled corpora and then is customized for downstream tasks by fine-tuning on a limited number of labeled samples \cite{han2021pre}.
Unlike TML and DL models that construct an AWI model from scratch, PTM can be directly used as a starting point.
Fig. \ref{fig:rq4-modelcategory} shows that two studies \cite{S44, S19} attempt PTMs for AWI. Specifically, Kharkar et al. \cite{S44} pre-train CodeBERTa \cite{codeberta} on a newly collected warning dataset and use a well-pre-trained CodeBERTa to classify targeted warnings. 
Also, they generate the warning-related source code recommendation to infer the legality of targeted warnings via GPT-C \cite{gpt-c}.
Similar to the usage of CodeBERTa in the work of Kharkar et al., Yedida et al. \cite{S19} train CodeBERT \cite{codebert} on the labeled warnings and use the well-trained CodeBERT for AWI.

As shown in Fig. \ref{fig:rq4-modelcategory}, 12 studies attempt multiple model categories for AWI. 
Specifically, 10 studies \cite{S5, S12, S21, S23, S24, S25, S26, S35, S53, S55} separately attempt TML and DL models for AWI. 
One study \cite{S44} separately uses TML model and PTMs for AWI. One study \cite{S19} uses TDP for AWI respectively.



\subsubsection{Learning category in AWI} 
Almost all primary studies, except for the work of Tu et al. \cite{S55}, \hl{perform AWI in a supervised learning manner}. That is, these studies train an AWI model in the labeled warnings and use this model for unlabeled warnings. 
By contrast, Tu et al. adopt semi-supervised learning for AWI. 
Specifically, they rely on an unsupervised learning technique called CLAMI \cite{nam2015clami} to assign pseudo-labels for warnings via clustering, select warning features via metric violation score, and determine warning instances for the AWI model training. However, CLAMI has many hyperparameters, which greatly affect AWI performance. To address this problem, they search for the optimal hyperparameters on a few labeled warnings. 
Under the optimal hyperparameters, they finally use the determined warning instances with the associated pseudo warning labels and the selected warning features to train a classifier for AWI.

\subsubsection{AWI model structure}
By analyzing the primary studies, the AWI model involves the base, ensemble, sequence, and multiple models-based structures. 

\begin{figure}
    \centering
    \subfloat[Model categories in AWI.]{
    \label{fig:rq4-modelcategory}
    \includegraphics[scale=0.495]{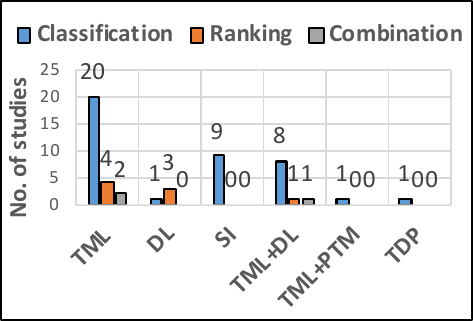}}
    \subfloat[AWI model structures.]{
    \label{fig:rq4-classifierstructure}
    \includegraphics[scale=0.495]{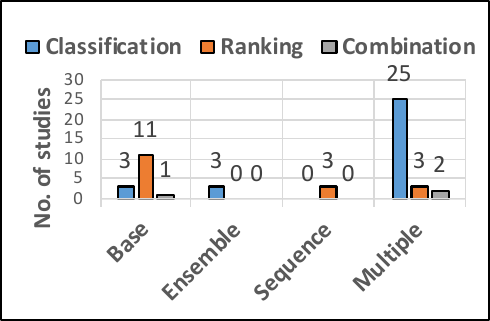}}
    \subfloat[AWI construction scenarios.]{
    \label{fig:rq4-constructionscenario}
    \includegraphics[scale=0.495]{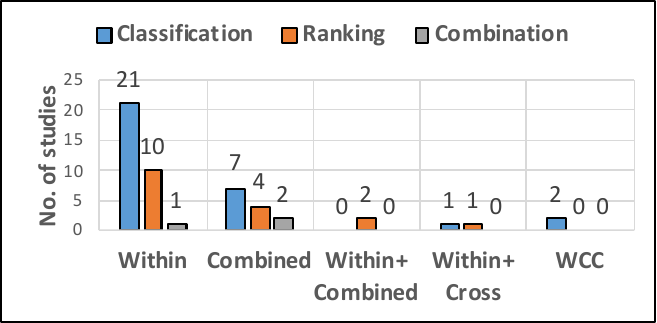}
    }
    \caption{Distribution of model categories in AWI, AWI model structures, and AWI construction scenarios in three categories of ML-based AWI approaches. SI is the statistical inference. TDP is TML+DL+PTM. WCC is to separately adopt within, combined, cross projects for construction scenarios for AWI in the same study.}
    \label{fig:rq4}
\end{figure}

\textbf{Base structure.} The base structure uses a single ML model for AWI. 
Fig. \ref{fig:rq4-classifierstructure} describes that 15 studies only rely on the base structure for AWI, where \hl{3} \cite{S7, S30, S32}, 11 \cite{S16, S33, S36, S37, S38, S39, S40, S41, S50, S51, S52}, and \hl{1} \cite{S8} studies fall into classification, ranking, and combination approaches respectively. 
The base structure is easy to implement and deals with massive warnings with high-dimensional features. However, the base structure could face the overfitting problem.

\textbf{Ensemble structure.} The ensemble structure relies on bagging, boosting, or stacking to combine multiple ML models for AWI. 
Different from the off-the-shelf ensemble learning models (e.g., RF), the ensemble structure borrows the core idea of ensemble learning and independently designs a new classifier/sorter by combining multiple ML models. 
Fig. \ref{fig:rq4-classifierstructure} shows that three studies \cite{S17, S46, S48} related to the classification approach adopt the ensemble structure for AWI.
For example, D2A \cite{S48} applies a soft-voting strategy to combine the results of four ML models for AWI.
Compared to the base structure, the ensemble structure can reduce the overfitting problem to some extent, obtain higher AWI performance, and be more stable and reliable. 
However, the ensemble structure is slower than the base structure when encountering massive warnings with high-dimensional features because each ML model in the ensemble structure needs to conduct an AWI model.

\textbf{Sequence structure.} The studies using the sequence structure divide a complex problem into sub-problems and sequentially solve each problem by using a proper ML model. 
Fig. \ref{fig:rq4-classifierstructure} shows that three studies \cite{S15, S27, S42} related to the ranking approach use the sequence structure for AWI. 
For example, given the control flow-based abstract syntax tree and program slice of warnings as input, Vu et al. \cite{S15} use CNN to encode these features, then use BiGRU to further learn the distribution of results encoded by CNN, and finally use dense layers to prioritize warnings. 
The sequence structure could identify more complex warnings while being slower than base and ensemble structures. 
Also, it requires further exploration about reasonably partitioning the AWI problem into sub-problems and \hl{deliberately} selecting proper ML models for sequentially solving each sub-problem.

\textbf{Multiple models-based structure.} The multiple models-based structure uses different ML models for AWI each time, thereby helping search for the optimal AWI performance. 
Fig. \ref{fig:rq4-classifierstructure} shows that the multiple models-based structure involves the most studies, where 25 \cite{S2, S3, S4, S6, S9, S10, S12, S13, S14, S18, S19, S20, S21, S24, S25, S26, S29, S34, S35, S44, S45, S49, S53, S54, S55}, \hl{3} \cite{S1, S5, S11}, and \hl{2} \cite{S23, S28} studies are related to classification, ranking, and combination approaches respectively.  
The multiple models-based structure can help investigate the performance differences of multiple ML models for the AWI task.
However, since the same task requires repeating multiple times, the multiple models-based structure is generally more time-consuming than the above three structures.

\subsubsection{AWI construction scenario} 
By analyzing the primary studies, the AWI construction scenario can be summarized into three categories, including within, combined, and cross projects. 

\textbf{Within project.} The within project \hl{is that} the warning dataset used for the model construction (i.e., the training set), and model evaluation (i.e., the test set) is from the same project. 
It indicates that the training and test sets have a similar distribution, thereby yielding superior AWI performance. However, the number of the training set is frequently insufficient.
As shown in Fig. \ref{fig:rq4-classifierstructure}, 63\% (32/51) of primary studies only use the within project as the AWI construction scenario. Of these, 21 \cite{S2, S3, S4, S6, S9, S13, S14, S17, S18, S19, S20, S21, S26, S29, S32, S45, S46, S48, S49, S53, S55}, 10 \cite{S1, S27, S33, S36, S38, S39, S42, S50, S51, S52}, and \hl{1} \cite{S28} studies fall into the classification, ranking, and combination approaches respectively. 

\textbf{Combined project.} The combined project merges warnings in multiple projects as a whole and partitions the merged warning dataset into the training and test sets.
Fig. \ref{fig:rq4-classifierstructure} shows that 13 studies separately adopt the combined project as the AWI construction scenario, where seven \cite{S7, S12, S30, S34, S35, S44, S54}, four \cite{S5, S11, S37, S40}, two \cite{S8, S23} studies are related to classification, ranking, and combination approaches respectively. 
The difference between within and combined project-based construction scenarios is that the warning dataset in the former is from the same project, and the warning dataset in the latter is merged from multiple projects. 
The combined project can alleviate the weakness of the within project to some extent\hl{,} while disrupting the original distribution of warning dataset in the within project.

\textbf{Cross project.} The cross project \hl{is} that the warning dataset used for the model construction (i.e., the training set) and model evaluation (i.e., the test set) is from different projects. 
It indicates that the training and test sets are heterogeneous.
Only two studies \cite{S10, S43} separately adopt the cross project as the AWI construction scenario.
The cross project can greatly mitigate the problem of warning dataset sparsity in the within project. 
However, due to the heterogeneous distribution in the training and test sets, it is more difficult to obtain the high AWI performance in the cross project than in the within project.

As shown in Fig. \ref{fig:rq4-classifierstructure}, six studies attempt multiple categories of AWI construction scenarios. 
Specifically, two \cite{S15, S16}, two \cite{S10, S41}, and two \cite{S24, S25} studies \hl{separately} adopt within and combined projects, within and cross projects, and WCC projects for AWI.

\begin{tcolorbox}[colback=gray!13, colframe=black, boxrule=0.3mm, boxsep= -0.1cm, middle=-0.1cm]
  \textbf{Summary RQ4}: The AWI model construction stage in the ML-based AWI approaches involves model category, learning category, AWI model structure, and AWI construction scenario.
  Specifically, ML models used for AWI are almost based on supervised learning, where the TML model occupies the majority.
  The multiple models-based structure is the most commonly used for AWI.
   The within project is the most prevalent AWI construction scenario. 
\end{tcolorbox}

\subsection{RQ5: How are ML-based AWI approaches evaluated?} \label{rq5}
The ML-based AWI model evaluation is critical for the practical applicability of ML techniques in AWI.
Based on the typical ML-based AWI workflow in Fig. \ref{fig:workflow}, 
the AWI model evaluation mainly involves two parts, including validation strategy and performance measure.
To present details in the AWI model evaluation stage, we analyze the techniques used in the two parts, along with associated strengths, weaknesses, and distribution across three categories of ML-based AWI approaches.

\subsubsection{Validation strategy in AWI} 
By analyzing primary studies, the validation strategy to evaluate ML-based AWI performance includes \emph{K}-fold, HoldOut, Leave \emph{P} Out, and rolling cross validation.

\begin{figure}
    \centering
    \subfloat[Validation strategies.]{
    \label{fig:rq5-validationstrategy}
    \includegraphics[scale=0.535]{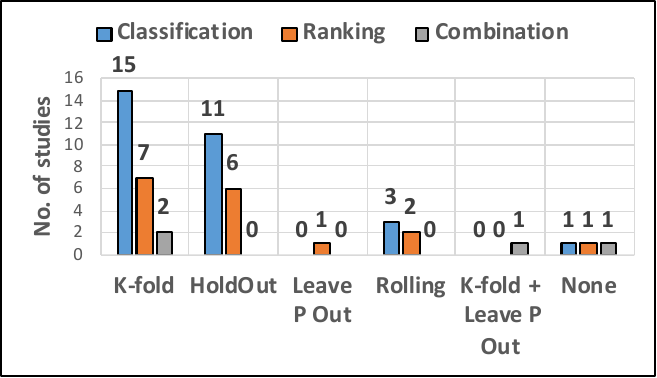}}
    \subfloat[Performance measures.]{
    \label{fig:rq5-performancemeasure}
    \includegraphics[scale=0.535]{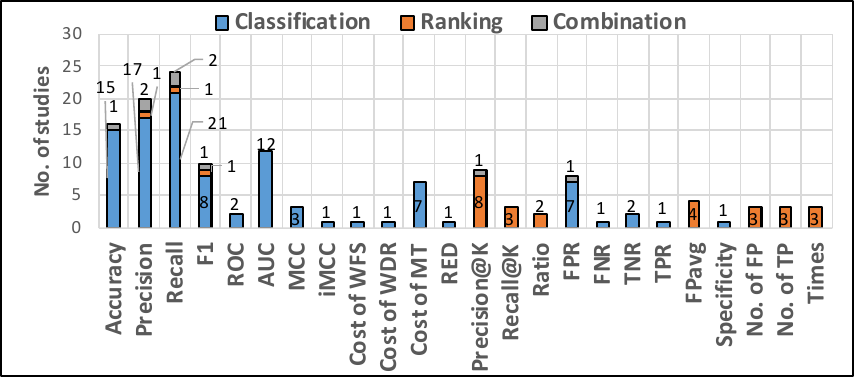}}
    \caption{Distribution of validation strategies and performance measures under the three categories of ML-based AWI approaches. MCC is Matthews Correlation Coefficient. iMCC \cite{S4} is the MCC difference between two approaches. Cost for WFS, WDR, and MT denote the time of warning feature selection, warning dataset rebalancing, and model training, respectively. RED \cite{S18} is the redundancy rate of warning features. Ratio \cite{S41, S42} is the performance ratio of the proposed ranking approach against the random ranking. FP$_{avg}$ is the average cumulative number of unactionable warnings before actionable warnings are found \cite{S1, S27, S41, S42}. Times \hl{are} the iterations for all discovered bugs in the ML-based AWI ranking approach.}
    \label{fig:rq5}
\end{figure}

\textbf{\emph{K}-fold cross validation}. K-fold cross validation divides the warning dataset into \emph{K} equally-sized folds. The ML model is trained and tested \emph{K} times, where a fold is used for the test set and the remaining \emph{K-1} folds are used for the training set each time. 
In general, \emph{K} is set to 10 \cite{S1, S2, S3, S4} or 5 \cite{S11, S15, S16, S24, S25, S44}.
Fig. \ref{fig:rq5-validationstrategy} shows that 24 studies adopt \emph{K}-fold cross validation, where 15 \cite{S2, S3, S4, S6, S7, S14, S17, S18, S24,S25, S35, S44, S45, S46, S49}, \hl{7} \cite{S1, S5, S11, S15, S16, S27, S42}, \hl{2} \cite{S23, S28} studies involve classification, ranking, and combination approaches respectively. 
In \emph{K}-fold cross validation, the entire warning dataset is used as the training and test sets, which can mitigate the estimation bias \cite{kohavi1995study}. 
However, for the imbalanced warning dataset, \emph{K}-fold cross validation may not ensure that each fold maintains the same class proportions as the whole warning dataset, thereby biasing the ML-based AWI performance \cite{follow}. 
Also, due to the iterative nature, \emph{K}-fold cross validation has high computational time.



\textbf{HoldOut cross validation.} HoldOut cross validation randomly divides the warning dataset into the training and sets. 
The common split ratios in HoldOut cross validation are 70/30 \cite{S29, S32} or 80/20 \cite{S19, S26, S32}. 
Fig. \ref{fig:rq5-validationstrategy} shows that 17 studies adopt HoldOut cross validation, where 11 \cite{S12, S13, S19, S21, S26, S29, S32, S34, S48, S54, S55} and \hl{6} \cite{S36, S37, S39, S41, S51, S52}) studies are related to classification and ranking approaches respectively. 
In comparison to \emph{K}-fold cross validation, HoldOut cross validation can finish the AWI model evaluation faster. However, HoldOut cross validation highly relies on the class distribution and split ratio, which could cause overfitting \cite{follow}.


\textbf{Leave \emph{P} Out cross validation.} In Leave \emph{P} Out cross validation, \emph{P} warnings are left as the test set, and the remaining warnings are used as the training set. This process is repeated for all possible combinations of leaving out \emph{P} samples from the warning dataset.
Only one study \cite{S40} related to the warning classification adopts Leave \emph{P} Out cross validation.
Similar to \emph{K}-fold cross validation, the entire warning dataset in Leave \emph{P} Out cross validation is used for the training and test sets, which can reduce the estimation bias \cite{kohavi1995study}.
Compared to HoldOut cross validation, Leave \emph{P} Out cross validation is more time-consuming when used for the AWI model evaluation.


\textbf{Rolling cross validation.} Rolling, aka time series-based, cross validation divides the warning dataset into the training and test sets in chronological order. 
For example, Wang et al. \cite{S10} build an AWI classifier in \hl{the warnings reported from} a prior revision and use \hl{the warnings reported from} a later revision to evaluate the performance of this classifier. 
Fig. \ref{fig:rq5-validationstrategy} shows that five studies adopt the rolling cross validation, where three \cite{S9, S10, S20} and two \cite{S33, S38} studies fall into classification and ranking approaches respectively. 
Rolling cross validation can reflect the real-world validation scenario, while being incompatible with the above three strategies due to time-sensitivity. 

As shown in Fig. \ref{fig:rq5-validationstrategy}, one study \cite{S30} separately attempts K-fold and HoldOut cross validation for AWI model evaluation. 
In addition, three studies \cite{S8, S50, S53} do not clearly mention the validation strategies for AWI model evaluation.

\subsubsection{Performance measure in AWI} 
As shown in Fig. \ref{fig:rq5-performancemeasure}, 47\% (24/51) of primary studies use Recall, followed by Precision (39\%) and Accuracy (31\%).
Precision@K and FP$_{avg}$ are the most frequently used by the ranking approach, which involve eight \cite{S5, S11, S15, S16, S33, S37, S50, S51} and four \cite{S1, S27, S41, S42} studies respectively.
Nine studies use Cost to evaluate the efficiency of sub-parts in the classification approach, including the warning feature selection \cite{S18}, warning dataset rebalancing \cite{S4}, and model training \cite{S24, S25, S26, S10, S20, S21, S32}.
In particular, one study \cite{S18} designs RED to evaluate the performance difference among warning feature selection techniques. 
One study \cite{S4} uses iMCC, derived from MCC, to evaluate the MCC difference between two ML-based AWI approaches.
Two studies \cite{S41, S42} adopt \hl{Ratio} to evaluate the performance difference between the proposed ranking approach and random ranking.

Since different studies conduct experimental evaluations on different warning datasets, we focus on analyzing the performance results in 10 studies with the same warning dataset. 
It is observed that on the same noisy and duplicate warning samples with data-leaked warning features, the AUC of eight studies \cite{S1, S3, S4, S10, S18, S20, S21, S55} almost falls into 70\%$\sim$90\%. 
However, on the same clean warning features and samples obtained via the manual filtering in the work of Kang et al. \cite{S9}, AUC \hl{generally drops to} 50\%$\sim$70\% \cite{S9, S18, S19}.
\hl{This} means \hl{that there is the} substantial room for AWI performance improvement.


\begin{tcolorbox}[colback=gray!13, colframe=black, boxrule=0.3mm, boxsep= -0.1cm, middle=-0.1cm]
  \textbf{Summary RQ5}: The AWI model evaluation stage in the ML-based AWI approaches contains the validation strategy and performance measure. 
  Specifically, \emph{K}-fold cross validation occupies the majority, followed by HoldOut and rolling cross validation.
  Recall, Precision, and Accuracy are the most commonly used performance measures in primary studies.
\end{tcolorbox}

%% file: src/5discussion.tex
\section{Future Research Directions} \label{discuss}
Based on \hl{the above analysis results in primary studies}, we highlight research directions for the future ML-based AWI field from the \hl{perspectives} of \hl{data improvement and model exploration} in the typical ML-based AWI workflow.

\subsection{Research Directions in Data Improvement} 
\textbf{(1) Collecting warning datasets with various SCAs and project development languages.} 
As shown in Section \ref{rq2}, most primary studies collect warning datasets from FindBugs and CppCheck as well as Java and C/C++ projects, respectively. 
However, few studies focus on SCAs (e.g., the symbolic execution-based SCA called Klocwork\footnote{https://www.perforce.com/products/klocwork}) and project development languages (e.g., Python) for the warning dataset acquisition. 
Also, the analysis results in Section \ref{rq2} indicate that due to SCAs with various techniques and project development languages, the acquired warning dataset is vastly diverse. 
Such diversity can help evaluate the generalizability of the ML-based AWI approach \cite{miningha, ViolationTracker}. 
In addition, it is observed that the warning datasets in only 17 primary studies are traceable. 
Specifically, the most commonly used and available real-world warning dataset, collected from FindBugs and Java projects, has been adopted by 10 studies \cite{S1, S3, S4, S9, S10, S18, S19, S20, S21, S55}. However, such a warning dataset contains mislabeled and duplicate samples \cite{S9}. 
The warning datasets used in two studies \cite{S24, S25} can be available. However, the warning datasets contain many synthetic warnings and only a small part of real-world warnings. 
In particular, there are only about 400 real-world warnings used for AWI in the study \cite{S24}. 
The remaining studies \cite{S2, S5, S36, S37, S52} either only disclose the real-world projects rather than the warning dataset corresponding with these projects, or collect warning datasets from the synthetic project source. 
Thus, it is essential to collect the diverse and real-world warning dataset and encourage disclosure of the collected warning dataset to stimulate the ML-based AWI process.

\textbf{(2) Improving the labeling strategy to construct more reliable warning datasets.} 
It is summarized in Section \ref{rq2} that the automatic and manual strategies are two main ways to label warnings.
The automatic labeling strategy, especially the closed warning-based heuristic \cite{S9}, is an important tactic to automatically construct a large-scale and real-world dataset.
However, such a heuristic is extremely sensitive to the warning-irrelevant source code changes, thereby causing mislabeled warnings \cite{S10}. 
Although some studies \cite{miningha, ViolationTracker, benchsource} are proposed to improve such a heuristic, there is still much improvement room for the warning labeling accuracy due to limitations of the warning-irrelevant source code discernment algorithm. 
It indicates that the current state-of-the-art heuristic is not still robust enough to label warnings. 
By contrast, solely performing the warning labeling via the manual strategy is not easy to scale and may be subject to the bias of an annotator \cite{S9}. 
Despite combining manual and automatic strategies for warning labeling in the primary studies, the current hybrid strategy mainly focuses on serialization (i.e., the manual and automatic strategies are used in sequential order to label warnings) rather than collaboration (i.e., the manual and automatic strategies are used interactively to label warnings). 
\hl{Such a serialization way} results in the inability to fully utilize the role of manual and automatic strategies in the warning labeling process.
Thus, we believe that it could be a promising labeling strategy via the human-machine collaboration to construct a reliable warning dataset. That is, the heuristic can gather different information (e.g., the source code revision message and the developer activity) to enrich warnings, thereby helping simplify the manual labeling process for annotators. 
In turn, the manual annotation could provide domain knowledge for complex warnings, thereby helping the heuristic obtain more precision warning labels.

\textbf{(3) Characterizing warnings with more fine-grained features for AWI.} 
First, Section \ref{rq3} shows five categories of warning features in the primary studies. 
A few studies \cite{S24, S25, S44} only mention that the sequential warning features can achieve superior performance in comparison to the characteristic-based and statistical warning features. 
However, it is not unclear \hl{which of the five warning feature categories in Section \ref{rq3}} shows the most powerful AWI ability. 
Further, although some studies (e.g., \cite{S10}) attempt different categories of warning features (e.g., characteristic-based, statistical, and content-based) for AWI, these studies are unaware of what combinations of different warning feature categories could maximize AWI performance.
Thus, it is necessary to conduct a comprehensive empirical evaluation on different categories of warning features and their combinations, thereby seeking out more precise and thorough warning features for AWI.
Second, in comparison to the other four categories, \hl{the structural category is the most expressive in the root cause of actionable and unactionable warnings} due to being able to grasp the intrinsic information of reported warnings (i.e., syntax and semantics) \cite{alexiadou2014syntax}.
In nine studies related to the structural category, warning features are mainly extracted by using the program slicing \cite{weiser1984program} to obtain the program dependency graph of the warning or using the def-use analysis \cite{defuse} to construct the datalog derivation graph of the warning. 
However, due to the limitations of program slicing and def-use analysis techniques \cite{weiser1984program, defuse, S39}, there is still warning-irrelevant or imprecision information in the program dependency graph and datalog derivation graph. 
Thus, it is essential to use more precise static analysis (e.g., SMT \cite{junker2012smt}) and more explicit dynamic execution tactics (e.g., fuzzing \cite{dynamic}) for acquiring sufficiently structural information but eliminating irrelevant information, thereby extracting more rigorous and complete warning features for AWI. 
Third, the SCA report is one of the important warning feature sources. The warning message (i.e., a warning characteristic in the SCA report) summarizes the basic warning information, which can provide auxiliary information for AWI \cite{rastkar2014automatic}. 
However, few studies consider the warning message for AWI. 
In the future, it could be a useful way to further enrich warnings for AWI by incorporating the warning message into warning features.





\textbf{(4) Employing feature selection techniques to enhance the AWI performance.} 
Through the fractal-based method \cite{levina2004maximum}, Yang et al. \cite{S21} experimentally prove that the original warning features in the work of Wang et al. \cite{S10} are inherently low-dimension.
Similarly, Ge et al. \cite{S18} observe that the original warning features \cite{S19} contain irrelevant and redundant features via PCA. 
Also, the evaluation in five studies \cite{S2, S6, S10, S14, S48} explicitly reveals that the warning features processed by feature selection techniques are more discriminative for AWI. 
In particular, instead of using the off-the-shelf feature selection techniques \cite{S2, S6, S10, S14, S48}, UNEASE\cite{S18} designs an unsupervised warning feature selection method for AWI, which can obtain the top-ranked AUC while maintaining low feature selection cost and feature redundancy rate. 
The above findings demonstrate that there could be irrelevance and redundancy in the original warning features, and the AWI performance can be further amplified after applying feature selection techniques to the original warning features.
However, most primary studies directly adopt the original warning features for AWI, while only 11 studies adopt feature selection techniques for AWI. 
Thus, researchers and practitioners should pay more attention to the role of feature selection techniques in AWI.
Further, it could be useful to employ feature selection techniques for enhancing AWI performance.

\textbf{(5) Selecting elaborately class rebalancing techniques to improve the AWI performance.} 
The warning dataset in the primary studies presents a prevailing phenomenon, i.e., class imbalance. 
Such a phenomenon limits the ML-based AWI performance \cite{S4, S48}.
However, most primary studies (\hl{i.e.,} 43) ignore the class imbalance when performing ML-based AWI. 
Only nearly 16\% (8/51) of primary studies attempt the class rebalancing technique to mitigate the class imbalance, thereby performing ML-based AWI. 
In particular, Ge et al. \cite{S4} investigate whether the off-the-shelf class rebalancing techniques can consistently improve the ML-based AWI performance. 
The experimental results describe that most class rebalancing techniques can significantly improve ML-based AWI performance. Surprisingly, a small part of class rebalancing techniques (e.g., KMeans-SMOTE) does not work for AWI in the imbalanced warning datasets. 
As such, researchers and practitioners should be concerned about the impact of class imbalance on AWI performance and further select elaborately the class rebalancing technique to improve AWI performance.

\subsection{Research Directions in Model Exploration}

\textbf{(1) Attempting semi-supervised or unsupervised learning for AWI.} 
The ML paradigm mainly contains supervised, semi-supervised, and unsupervised learning. 
As shown in Section \ref{rq4}, \hl{almost all} ML-based AWI approaches in the primary studies follow supervised learning for AWI, which often requires massive labeled warnings to achieve high AWI performance. 
However, as shown in the warning dataset preparation of Section \ref{rq2}, massive reliable and labeled warnings are often hard to gather quickly in practice \cite{ViolationTracker, miningha, benchsource, S9}. 
By contrast, semi-supervised learning can use a small portion of labeled samples and lots of unlabeled samples to train a predictive model, and unsupervised learning can identify patterns in unlabeled samples.
Particularly, Tu et al. \cite{S55} is the first to adopt semi-supervised learning for AWI and achieve superior AWI performance.
Thus, semi-supervised or unsupervised learning is strongly recommended for AWI, so as to help reduce the number of labeled warnings as many as possible while maintaining high AWI performance.

\textbf{(2) Exploring AWI via large language models.} 
Large Language Models (LLMs) are PTMs with larger-scale corpora and more training parameters. 
Similar to PTMs, LLMs can directly capture the rich sequential, syntactic, and semantic information from a given source code snippet. 
This is because the well-designed attention mechanism in the transformer structure of LLMs has the ability to localize areas of interest, thereby helping explain the factors contributing to specific tasks (e.g., vulnerability detection) and effectively boosting their performance \cite{structcode2, structcode1, sws}.
Besides, pre-training (i.e., training on large-scale and unlabeled corpora) and fine-tuning (i.e., training on a few labeled samples related to the downstream task) are proved to play critical roles in LLMs \cite{ge2024pre}.
Currently, LLMs have already exhibited tremendous potential in various software engineering tasks \cite{liu2023pre}.
However, only two studies \cite{S44, S19} attempt PTMs for AWI.
Despite revealing higher AWI performance in PTMs compared to TML and DL models, Kharkar et al. \cite{S44} only conduct a preliminary study for AWI on two obsolete PTMs (i.e., CodeBERTa and GPT-C). 
It indicates that the ability of current LLMs (e.g., ChatGPT4\footnote{https://chat.openai.com/}) has not been fully investigated in the ML-based AWI community.
As many state-of-the-art LLMs are proposed and released, an immediate direction is to conduct systematic experiments to understand the merits and shortcomings of LLMs in AWI, thereby exploring how to use LLMs for enhancing AWI. 
In addition, it could be an effective way to enlarge the benefits of pre-training (e.g., developing domain-specific LLMs by pre-training the AWI-related task) and fine-tuning (e.g., enhancing AWI performance by fine-tuning LLMs on massive warning datasets) when applying LLMs to AWI.



\textbf{(3) Designing the customized AWI approaches for different categories of warnings.} 
Different categories of warnings have different characteristics.
For example, warnings with cross-site scripting (XSS) generally contain URI schema, host name, or port number. 
Once an XSS vulnerability in a project is attacked, the privacy information of this project is leaked. 
By contrast, warnings with bad practices are related to the source code writing standards, which do not necessarily cause software defects. 
This indicates that an ML-based AWI approach used to identify one category of warnings could not perfectly fit the other category of warnings. 
However, nearly 65\% (33/51) of primary studies ignore the impact of warning categories on AWI, i.e., adopt a designed ML-based AWI approach to sweepingly identify all categories of warnings reported by a SCA.
Although the remaining 18 primary studies \cite{S5, S7, S11, S13, S14, S15, S16, S29, S30, S35, S36, S37, S38, S39, S40, S41, S44, S52} design the corresponding ML-based AWI approaches to identify one or several specific warning categories (e.g., datatrace \cite{S37}), these studies ignore the exploration about the performance difference of the designed ML-based AWI approach on different categories of warnings. 
In future work, it is necessary to make \hl{the} cons and pros of different ML-based AWI approaches on different categories of warnings, thereby helping select a proper ML-based AWI approach for a given warning category. 
Further, it is strongly recommended to design the customized ML-based AWI approach for different categories of warnings, thereby contrapuntally enhancing AWI performance.

\textbf{(4) Conducting the practical AWI model evaluation via rolling cross validation.} 
As shown in Section \ref{rq5}, the common strategy used for the ML-based AWI model evaluation is \emph{K}-fold and HoldOut cross validation, which occupy 80\% of primary studies. 
The two validation strategies, especially for \emph{K}-fold cross validation, often make the sample distribution between the training and test sets consistent \cite{ge2021impact}. 
Conversely, in the real-world scenario, warnings in the training set should be historically prior to that of the test set. 
It indicates that warnings in the test set may be similar to warnings in the training set, contain warnings that are unknown to the training set, or even be entirely different from warnings in the training set. 
Thus, the two validation strategies ignore the warning timelines in the real-world scenario, which could result in exaggerating the performance of ML-based AWI approaches \cite{ge2021impact, pendlebury2019tesseract}.
The rolling cross validation, dividing the warning dataset into the training and test sets in chronological order, can exactly satisfy the ML-based AWI model evaluation requirement in the real-world scenario. 
However, only five primary studies \cite{S9, S10, S20, S33, S38} consider the rolling cross validation as the ML-based AWI validation strategy. 
It indicates that the performance of ML-based AWI approaches in most primary studies could be biased in the real-world scenario. 
Thus, we recommend adopting the rolling cross validation to conduct the practical AWI model evaluation, thereby helping reveal the ML-based AWI performance in the real-world scenario.

%% file: src/6threat.tex
\section{Threats to Validity} \label{threat} 
\textbf{External.} The threat to external validity concerns whether the findings of our survey are generalizable beyond studies in the entire ML-based AWI population.
Since all primary studies are collected from the entire ML-based AWI population, this threat is not applicable.

\textbf{Internal.} The threat to internal validity concerns the problems of data extraction consistency and correctness. 
To alleviate the above problems, we conduct a pilot study and double manual verification in the process of data extraction (see in Section \ref{collectpaper}). 
   

\textbf{Construct.} The threat to construct validity concerns whether our collected primary studies are related to ML-based AWI or miss relevant ML-based AWI studies. 
To ensure the relevance and completeness of the collected primary studies, we first inherit and extend the search keywords from the latest AWI studies \cite{S31}. Besides, we design the selection criteria based on a pilot search and perform the selection of primary studies based on the 2-pass review. 
Moreover, we conduct the snowballing to help identify relevant studies that could be missed by the keywords-based search. 
Thus, we believe that this threat of construct validity can be mitigated in our survey.

%% file: src/7conclusion.tex
\section{Conclusion} \label{conclusion} 
In this paper, we conduct a comprehensive survey related to the ML-based AWI approach.
We first perform a meticulous survey methodology to collect ML-based AWI studies in five digital libraries from 2000 to 2023, thereby obtaining 51 primary studies. 
After that, we describe a typical ML-based AWI workflow, including the warning dataset preparation, preprocessing, AWI model construction, and evaluation. 
By following such a workflow, we rely on the warning output format to classify ML-based AWI approaches into three categories, involving classification, ranking, and combination approaches. 
Besides, we analyze the techniques used in each stage of such a typical workflow by discussing their strengths and weaknesses as well as presenting their distribution under the three ML-based AWI approach categories.
Finally, we rely on the analysis results to highlight practical research directions for the ML-based AWI community.

\section*{Acknowledgements} We would like to thank the anonymous reviewers for their insightful comments. The work is partly supported by the National Natural Science Foundation of China (61932012, 62372228, 62141215) and the Program of China Scholarship Council (Grant No. 202306190140).
